\documentclass[12pt]{article}
\usepackage[dvipdfmx]{graphicx}
\usepackage[version=3]{mhchem}

\usepackage{dcolumn}

\usepackage{afterpage}
\usepackage{amsmath}
\usepackage{amssymb}
\usepackage{amsthm}
\usepackage{arydshln}
\usepackage{ascmac}
\usepackage{bm}
\usepackage{booktabs}
\usepackage{caption}
\usepackage{cases}
\usepackage{color}
\usepackage{comment}
\usepackage{float}
\usepackage{floatpag}
\usepackage{longtable}
\usepackage{lscape}
\usepackage{multirow}
\usepackage{mathrsfs}
\usepackage{pdfpages}
\usepackage{rotating}
\usepackage{setspace}
\usepackage[subrefformat=parens]{subcaption}
\usepackage{tabularx}
\usepackage{url}
\usepackage{siunitx}

\usepackage[square, sort, comma, numbers]{natbib}
\bibpunct[, ]{(}{)}{;}{a}{}{,}

\newcommand{\ie}{i.e.}

\newcommand{\vx}{\mathbf{x}}

\newcommand{\vdelta}{\bm{\delta}}

\newcommand{\WWII}{\mbox{WWI\hspace{-.05em}I }}
\newcommand{\wwii}{\mbox{WWI\hspace{-.05em}I}}

\graphicspath{{./figure/}}

  

\begin{document}

\title{Far-reaching effects of bombing on fertility in mid-20th century Japan}

\author{
Tatsuki Inoue\thanks{School of Commerce, Meiji University, Faculty Office Building, 1-1, Kanda-Surugadai, Chiyoda-ku, Tokyo, 101-8301, Japan (E-mail: tatsukiinoue@meiji.ac.jp).}, 
Erika Igarashi\thanks{Institute for Advanced Study, Hitotsubashi University, 708, Faculty Building 2, 2-1, Naka, Kunitachi, Tokyo, 186-8601, Japan (E-mail: igarashi.erika715@r.hit-u.ac.jp).}
}

\date{}
\maketitle
\begin{abstract}
\noindent 
This study explores the indirect impact of war damage on postwar fertility, with a specific focus on Japan's air raids during World War I\hspace{-.05em}I.
Using 1935 and 1947 Kinki region town/village data and city air raid damage information, we explored the ``far-reaching effects'' on fertility rates in nearby undamaged areas.
Our fixed-effects model estimates show that air raids influenced postwar fertility within a 15-kilometer radius of the bombed cities. These impacts varied based on bombing intensity, showing both positive and negative effects.
Moreover, a deeper analysis, using the Allied flight path as a natural experiment, indicates that air raid threats and associated fear led to increased postwar fertility, even in undamaged areas.
This study highlights the relatively unexplored indirect consequences of war on fertility rates in neighboring regions and significantly contributes to the literature on the relationship between wars and fertility.

\bigskip

\noindent\textbf{Keywords:}
Air raids, Bombing; Fear; Fertility; World War I\hspace{-.1em}I

\bigskip


\end{abstract}

\newpage
\section{Introduction} \label{sec:intro}

Despite international peacekeeping efforts, wars and conflicts persist.
In 2020, there were 56 active conflicts worldwide \citep{Strand2021}.
More recently, the military crash between Russia and Ukraine in 2022 attracted global attention.
Given the current situation, it is increasingly essential to gain multifaceted foresight into postwar societies and to deepen our understanding of war.%
\footnote{The most recent economic research delving into postwar changes includes \citet{Boehnke2022, Padilla-Romo2023, Koenig2023, Akresh2023, Bose2022, Korovkin2023, Prem2023, Beaman2022, Vlachos2022}.
In addition, the origins of and preventive measures against wars and conflicts have been examined \citep{Narciso2023, Gallea2023, Mueller2022, Castells-Quintana2022, Rexer2022, Mayoral2022}.}
As seen in the mid-twentieth century baby boom, postwar fertility rate changes can profoundly impact society.
This study investigates the effects of war damage on postwar fertility from a new perspective by utilizing Japan's experience with air raids during World War \mbox{I\hspace{-.05em}I}.
We examine the far-reaching effects, which are the indirect effects of war damage on the fertility rate, in undamaged areas near war-damaged ones.

Japan offers an ideal setting for examining far-reaching effects.
Despite intensified air raids on cities in early 1945, towns and villages mostly escaped direct damage, as the Allied Forces focused solely on bombing cities.
Furthermore, city damage varied significantly, with the Kinki region having both war-damaged and non-damaged cities.%
\footnote{The Kinki region is located roughly in the center (slightly to the west) of Japan (see Online Appendix A).
This study treats Mie, Shiga, Kyoto, Osaka, Hyogo, Nara, and Wakayama Prefectures as the Kinki region according to the general classification.}
When controlling for relevant variables, city bombing damage is considered independent of fertility determinants in the surrounding rural areas.
Thus, we capture the far-reaching effects by focusing on fertility changes in towns and villages near bombed cities.
Japan's experience provides further helpful situations.
Allied aggression primarily involved aerial bombing, given Japan's geography.
These cities were devastated by intense air raids over a relatively brief period.
Furthermore, the Japanese government's covert information controls effectively limited the spread of the bombing's indirect effects.

Our dataset includes town- and village-level crude birth rates, a well-established measure of fertility in the historical context, in the Kinki region in 1935 and 1947, along with air raid damage information for cities during \wwii.
To eliminate the impact of direct damage and to focus on far-reaching effects, we excluded cities from our observations.
Comprehensive regional-level data helped resolve the potential issue of sample selection bias.
The data sources for this study are population censuses and the most reliable reports of air raid damage surveyed by the Economic Stabilization Board.
As an identification strategy, this study employs a fixed-effects model with the essence of a difference-in-differences approach, using the death toll from bombings in nearby cities as the treatment to represent the scale of the air raids.
Consequently, our estimates avoid the influence of war-related macro shocks and time-constant characteristics.

To identify some of the mechanisms behind the far-reaching effects, this study leverages the Allied flight path to Japan as a natural experiment.
The bombers of the Allied Forces only flew from the Mariana Islands, located southeast of Japan, due to flight distance limitations.
Consequently, southeastern towns and villages in bombed cities were exposed to stronger fear, a potential factor, than those in other directions.
However, other potential mechanisms are independent of the orientation of the cities.
Therefore, we can capture far-reaching effects through fear by investigating the differences in the effects between the southeast and other directions.
 
The estimation results suggest that air raids during \WWII affected the postwar fertility rate in towns and villages within at least 15 km.
While large-scale air raids positively affected postwar fertility rates, small-scale air raids had a negative impact.
A back-of-the-envelope calculation indicates that aerial bombings increased the crude birth rate by 3.90\% in severely damaged cases and decreased it by 4.74\% in other cases.
Moreover, the quasi-experimental approach demonstrates that the far-reaching effects of bombings are negative in municipalities located non-southeast of the nearest bombed city but are more positive in the southeast, especially in the severely damaged group.
These results suggest that the intense fear caused by air raids led to postwar increases in the fertility rate, while the negative effects were likely attributable to other factors, such as the enhancement of national morale.

The contributions of this study are as follows:
We discovered that air raids influenced postwar fertility, even in nearby areas that were not directly damaged.
While numerous studies have examined war's impact on fertility, focusing on macro and direct effects, our research explores the indirect effects on neighboring regions.%
\footnote{As for macro shocks, there is literature on the mobilization and demobilization \citep{Van_Bavel2013}, absence of men during wartime \citep{Woldemicael2008}, scarcity of men in the postwar period \citep{Brainerd2017, Bethmann2012, Kraehnert2019, Jayaraman2009, Kesternich2020}, gender imbalance in the marriage market \citep{OgasawaraKomura2021}, changes in the demand for female labor \citep{Doepke2015, Brodeur2022}, high relative income \citep{Easterlin1961, Hill2015}, decline in educational level \citep{Islam2016, Urdal2013}, economic difficulty and health deterioration \citep{Lindstrom1999, Kidane1989, Sobotka2011}, and political instability \citep{Williams2012}.
These shocks have broad impacts on participating countries or regions but are independent of individual-level war damage from direct attacks.
Studies on individual shocks examine child death \citep{Kraehnert2019, Heuveline2007}, sibling death \citep{Kraehnert2019}, an increase in child mortality risk \citep{Verwimp2005, Lindstrom1999}, and the threat of harm \citep{Williams2012}.}
However, the indirect effects of war damage on fertility rates in neighboring areas remain understudied.
\citet{Agadjanian2002} found a remarkable postwar rebound in fertility in the capital city of Angola, despite the minimal direct impact of the war, in response to the politico-military and economic changes.
However, this impact differs from the far-reaching effects we have focused on, which depend on distance and damage extent near affected areas.
Thus, this study adds a novel perspective to previous research on postwar birth rate changes.
Additionally, our findings imply that comparisons between damaged and non-damaged areas nearby may overestimate or underestimate the war's impact.

Moreover, this study presents new evidence contributing to the literature on short-term fertility changes in response to destructive shocks, such as natural disasters and terrorism \citep[for example,][]{Rodgers2005, Nobles2015, Cohan2002, Frankenberg2014, Kidane1989, Nandi2018}.
These previous studies have examined the impact of shocks and potential channels.
We have discussed how air raids could influence postwar birth rate in areas outside the bombed zones and have demonstrated that fertility increases in response to severe threats and fear.
Given the common mechanisms impacting fertility, our study elucidates the effects of different shocks on fertility.

This study also contributes to the literature on the baby boom in a defeated nation.
Unlike various explanations for the United States, Japan's fertility rate surge is often attributed to large-scale demobilization \citep{vsej1955}.%
\footnote{For accounts for the U.S. baby boom, see \citet{Easterlin1961, Hill2015, Doepke2015, Greenwood2005, Albanesi2014, Bailey2011}.}
While recent literature has focused on the impact of gender imbalance on the marriage market \citep{OgasawaraKomura2021} and the restoration of peace due to the return to normalcy \citep{Igarashi2022}, no studies have addressed the heterogeneity in the baby boom within the country.
Although not our study's main focus, we suggest air raids may have contributed to mid-twentieth-century fertility variations.

The remainder of this paper is organized as follows:
Section~\ref{sec:back} provides the historical background of this study.
Section~\ref{sec:emp} elaborates on the potential mechanisms, our data, and the baseline identification strategy used to investigate the far-reaching effects on postwar fertility.
The estimation results are presented in Section~\ref{sec:res}.
Finally, Section~\ref{sec:con} concludes the paper.

\section{Background} \label{sec:back}

\subsection{Air raids on cities} \label{sec:bar}
Starting with the first bombardment of the four mainland cities in April 1942, the Allied Forces launched extensive bombing campaigns against Japanese cities until August 1945, when Japan surrendered.
Between 1942 and 1944, they predominantly targeted munition factories, such as aircraft manufacturing plants and military arsenals.
However, hit rates remained low due to the relatively small size of the factories and the influence of weather conditions.
Hence, the Allied Forces shifted their bombing strategies in January 1945.
They believed that the domestic home industry was essential for Japan to continue the war and began bombarding all cities, including private houses and small factories engaged in the cottage industry.
The large cities designated as ``Selected Urban Industrial Concentrations'' were extensively destroyed by June 1945.
Subsequently, even small and medium-sized cities experienced severe air raids \citep{Okuzumi2006}.

The Kinki region had been consistently targeted by aerial bombardment owing to the presence of the Hanshin Industrial Zone, a major industrial area in Japan \citep{Tsujikawa1992}.
Osaka, the second-largest city after Tokyo, suffered significant infrastructure and human losses.
Following nighttime attacks that resulted in approximately 200 casualties in January and February 1945, the Osaka Air Raid on March 13 obliterated over 130,000 buildings and resulted in more than 3,000 fatalities.
Ultimately, Osaka endured 33 aerial bombings by the end of the war.
After the destruction of the Hanshin Industrial Area, the Allied Forces carried out precision bombings on military factories in other cities, such as Akashi, Nishinomiya, and Himeji.%
\footnote{The main targets included military-related facilities of large companies, such as the Kawasaki Aircraft Company, Kawanishi Aircraft Company, Toyo Boseki Company, Dai-Nihon Spinning Company, and Dai-Nihon Celluloid Company.
The bombing of the Kawanishi Aircraft plant in Akashi resulted in the casualties of 74 employees and approximately 200 civilians in the vicinity of the plant.}
The principal facilities were destroyed by June, after which the bombing targets shifted to residential sections in small- and medium-sized cities \citep{Kimimoto1980}.

\begin{figure}[!t]

\begin{minipage}{0.5\textwidth}
\centering
\includegraphics[keepaspectratio, width=\textwidth]{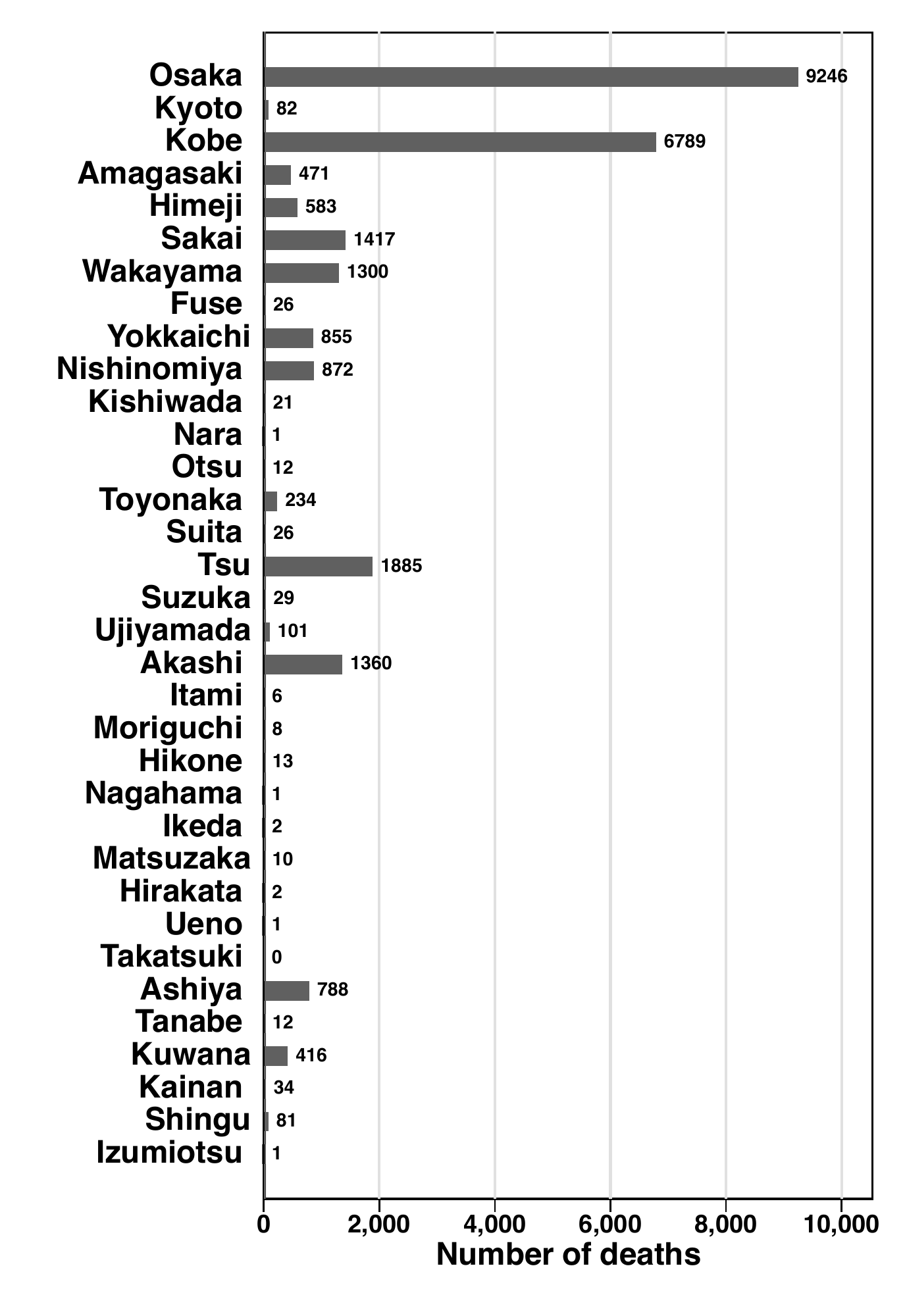}
\subcaption{Number of deaths}
\label{fig:vic2}
\end{minipage}
\begin{minipage}{0.5\textwidth}
\centering
\includegraphics[keepaspectratio, width=\textwidth]{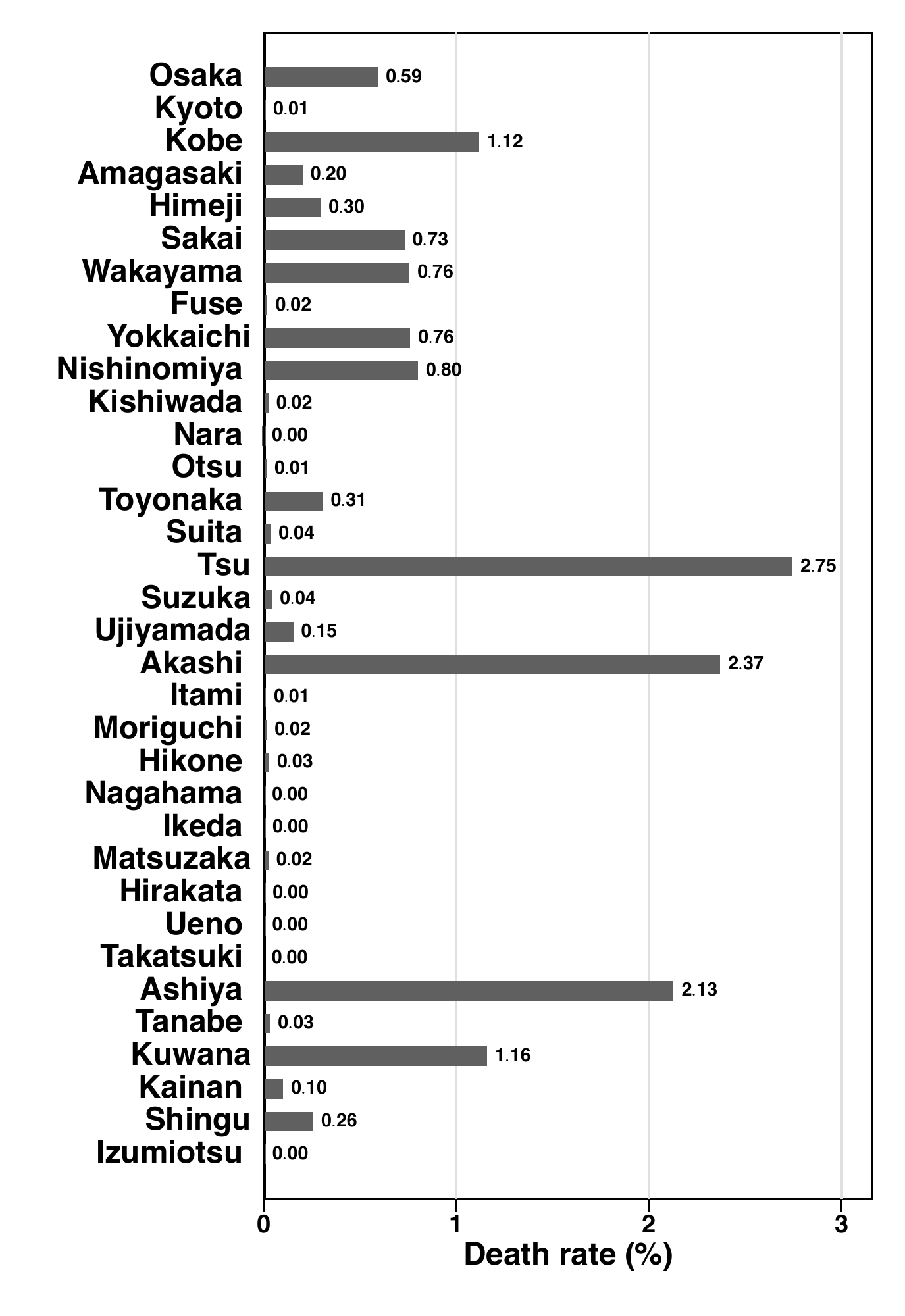}
\subcaption{Death rate (\%)}
\label{fig:vic3}
\end{minipage}

\caption{Human casualties due to air raids in cities in the Kinki region}
\label{fig:vic}

\begin{minipage}{156mm}
\scriptsize{
Notes:
The number of deaths from air raids comes from a survey in May 1948, and the death rates were calculated based on the population in 1947.

Sources: \citet{esb1949} and \citet{census1947pp}.
}
\end{minipage}

\end{figure}

Figure~\ref{fig:vic} shows human casualties in cities in the Kinki region, sorted by population in 1947.
Population size was a crucial factor in determining bombing targets, but it did not necessarily determine the scale of casualties.%
\footnote{The Allied Forces typically targeted cities from the list titled ``Attacks on Small Urban Industrial Areas,'' which ranked cities by population based on the 1940 census.
Despite the title, this list included 180 cities, including major urban centers such as Tokyo and Osaka.}
While Kyoto, with the second-largest population in Kinki, experienced relatively few casualties, Tsu, Akashi, Ashiya, and Kuwana suffered severe damage despite their relatively small populations.
Some cities avoided severe air raids due to external factors: Kyoto was excluded as a target by the directive of the Allied Command, Ikeda was mistakenly believed to be outside the flying range of bombers, and the geographical features of Fuse and Suita were unfavorable for aircraft radar \citep{Okuzumi2006}.%
\footnote{The reason Kyoto was spared from severe air raids, despite its large population, remains unclear due to insufficient evidence. Potential explanations include an intent to preserve its cultural and historical heritage, such as ancient temples, or consideration as a potential target city for the atomic bombing \citep[p. 310]{Kimimoto1980}.}

Air raids did not directly target people living outside cities, but they could have instilled terror in them.
The Allied Forces primarily conducted nighttime attacks to minimize encounters with Japanese forces \citep{Tokyo1973}. 
As suburban residents returned home at night, even if they worked in nearby cities during the daytime, they could avoid direct damage from air raids. 
However, those who lived near a bombed city may have experienced intense fear at the sight of aerial bombardment and a burning city.%
\footnote{According to \citet{Yokoyama2001}, eyewitness accounts indicated that the air raids on Sakai City and Wakayama City were visible from approximately 20 km away.}
A resident of a suburban area described this fear as follows:
\begin{quote}
In general, the people became terrified of the war, and the number of those continued to grow who stored their household baggage in air-raid shelters or intrusted [sic] it to relatives and friends in the far mountain regions. Especially just before war's end, air-raid warnings sounded morning, noon, and night, and there were few people who slept soundly for listening to the air-raid reports over the radio. \citep[p. 35]{USSBS}
\end{quote}

Meanwhile, people could only determine the exact extent of damage through visual confirmation.
This occurred because the Imperial General Headquarters underreported the extensive damage caused by air raids to the public through radio and newspapers, even though the damage was substantial.
Consequently, people in distant areas believed that the air raids were insignificant and, in some cases, remained unaware that aerial bombing had taken place. 
Despite the tremendous loss of approximately 100 thousand lives in the Great Tokyo Air Raid on March 10, 1945, the inhabitants of Osaka, located 400 km away from Tokyo, continued with their lives as usual, harboring little fear of air raids \citep{Koyama1985}.

\subsection{Relocation of the population during and after the war} \label{sec:btrp}

\subsubsection{Evacuation} \label{sec:bev}

The government enacted the Air Defense Law to prevent citizens from leaving cities and to compel them to participate in firefighting efforts during aerial bombings.%
\footnote{This law, initially established in October 1937 to mitigate the damages from air raids, underwent revisions in response to changing wartime circumstances.
In November 1941, it incorporated measures anticipating air raids, including the prohibition of evacuation from cities, the obligation to engage in emergency firefighting during air raids, and the imposition of stricter penalties.}
The prohibition on leaving cities was enforced through the mutual monitoring of neighborhood groups known as \textit{tonarigumi}, and those who violated this rule faced imprisonment or fines.
Additionally, local authorities frequently employed threats, such as removing individuals from the community register and discontinuing rations, to dissuade residents from attempting to seek refuge \citep[p. 12]{Mizushima2014}.
Being labeled as unpatriotic for abandoning civic duties also functioned as a form of social censure, further discouraging escape.

Some residents were granted exceptions and allowed to evacuate from cities. This included individuals who lived in or worked in buildings scheduled for demolition as part of building evacuations, victims of air raids, and people residing in cities where evacuation was recommended under the Urban Evacuation Implementation Guidelines established in December 1943.%
\footnote{Regarding building evacuations, approximately 610,000 buildings were demolished to create fire-blocking belts in 279 cities during \wwii.
Following the Osaka Air Raids, priority was given to the victims of air raids.
These individuals were directed to designated evacuation sites located in neighboring prefectures, such as Nara and Shiga, rather than to nearby towns and villages \citep{Japizoku2002}.
}
These guidelines were applied to 11 cities, including Osaka, Kobe, and Amagasaki in the Kinki region.
In March 1944, the government strongly recommended the evacuation of specific groups in these cities, such as older individuals (65 years and older), young children (elementary school students and younger), pregnant women, and individuals with illnesses, as they could hinder firefighting efforts during air raids \citep[p. 73]{Mizushima2014}.

However, during the wartime period, all evacuations required a license issued by the city office, limiting people's freedom to relocate.
Moreover, those who wished to evacuate were required to find an evacuation destination on their own, usually relying on relatives living in rural areas.
Furthermore, with the growing demand for evacuation, city offices ceased issuing licenses to civil servants, factory workers, and other essential workers who were crucial to maintaining city functions, even in cities where evacuation was recommended.
Consequently, only a limited portion of the population could evacuate from these cities.

\subsubsection{Demobilization and repatriation} \label{sec:bdr}

In the postwar period, numerous demobilized soldiers and Japanese residents returned from overseas territories, including Manchuria, Taiwan, and the Korean Peninsula.
By the end of 1949, the number of returnees had exceeded six million \citep[p. 12]{rra1950}.%
\footnote{Nevertheless, an estimated 376,929 people were still left behind at this time.
These unrepatriated people decreased to 63,042 by 1955 \citep[p. 162]{rra1963}.}
Upon their return, these individuals usually went back to their hometowns or the homes of relatives.
Limited available data suggests that demobilized and repatriated people represented approximately 2--4\% of the population in cities and 7\% in counties across the Kinki region.%
\footnote{In Shiga Prefecture, they accounted for 7\% of the population in both Otsu City and Gamo County and 3--4\% in other cities and counties as of October 1947 \citep[p. 39]{shiga1948}.
In Hyogo Prefecture, as of March 1949, the percentage was only approximately 2\% across various cities and counties, with the exception of 7\% \citep[pp. 70--71]{hyogo1950}.}

Prefectural governments encouraged returnees whose hometowns were destroyed by air raids and who had no relatives to support to move to temporary shelters or settlements and engage in agriculture \citep[pp. 89--90]{rra1950}.
For instance, in Wakayama Prefecture, 669 households settled in designated areas and cultivated their land between 1945 and 1947 \citep[p. 114]{wakayama1949}.
Nara Prefecture also promoted this welfare project, receiving 334 and 114 families in 1946 and 1947, respectively, \citep[p. 137]{nara1949}.
In 1947, the government of Shiga Prefecture authorized the settlement of 569 households to cultivate agricultural land \citep[p. 86]{shiga1948}.
According to \citet[pp. 162--163]{Yasuoka2014}, those who arrived in the postwar period were more likely to settle compared to those who came for evacuation during the war, at least in the case of Kyoto Prefecture.

In addition to the utilization of uncultivated land in rural areas, former military lands that the GHQ had not confiscated in cities were also converted into settlements.
Furthermore, prefectural governments allocated people to settlements in groups of 10–-100, ensuring that the population did not concentrate in one place despite the significant influx of returnees.

\subsubsection{Labor mobilization and transfer} \label{sec:blm}

In addition to evacuation and return, labor mobilization could have led to population relocation.
Owing to the shortage of male labor during wartime, the government forcibly mobilized unmarried women and workers from non-essential industries into war-related sectors.
Consequently, the population may have temporarily relocated, while the evacuation of workers was restricted.
Nevertheless, the impact on rural areas was likely limited because the agricultural sector was mostly excluded from mobilization to ensure an adequate food supply.%
\footnote{Male agricultural workers were also mobilized into military industries in the early stages of the war, when there was still spare capacity to produce food.
Consequently, in conjunction with conscription, the proportion of female labor increased in the agricultural sector during the war.}

\begin{figure}[t]
 \centering
 \includegraphics[keepaspectratio, scale=1.0] {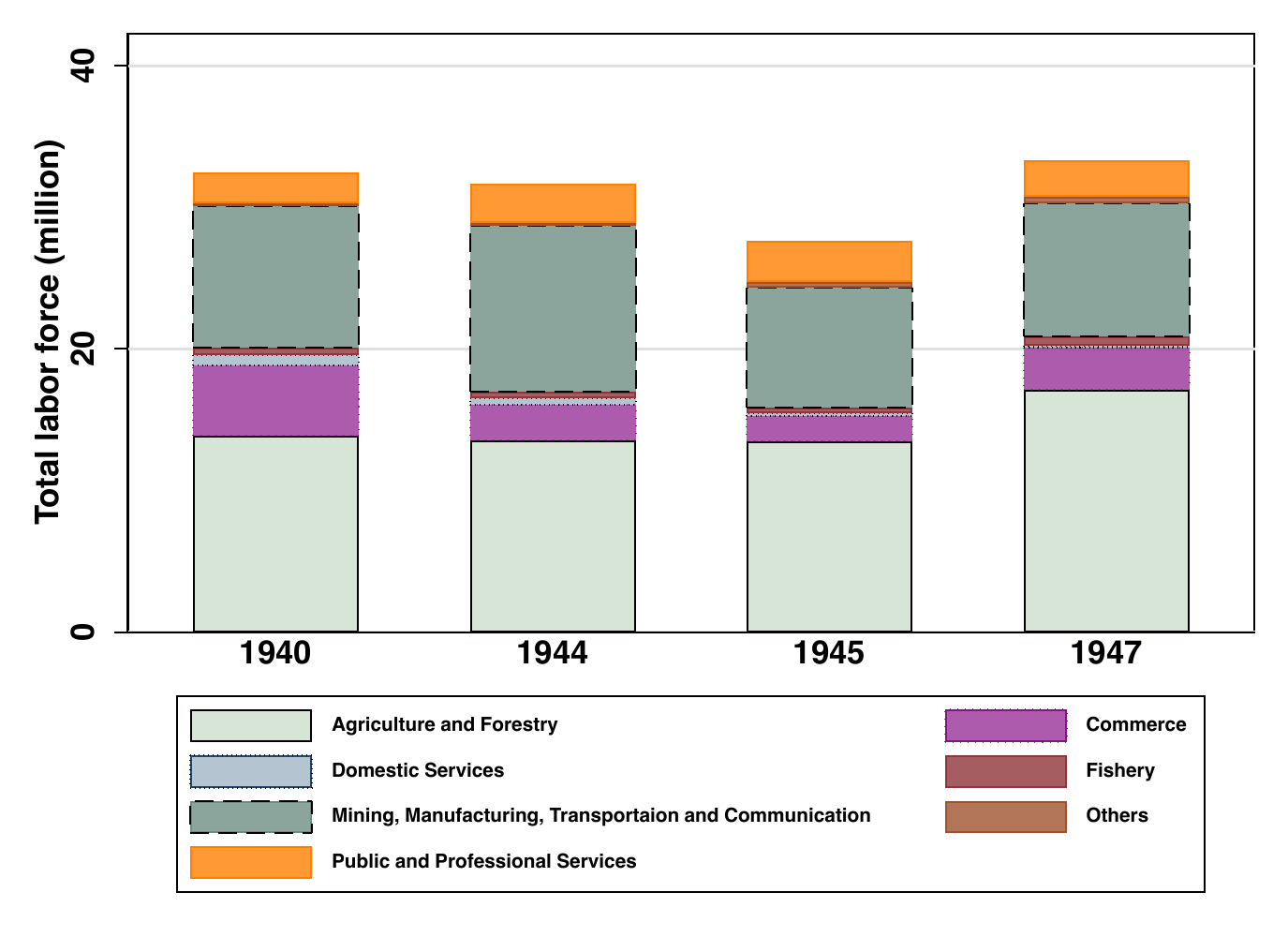}
 \caption{Trends in labor force by industry}
 \label{fig:lf}
 {\scriptsize
  \begin{minipage}{156mm}
  Notes:
  The labor forces were based on surveys in October 1940, February 1944, May 1945, and October 1947.
  Sources: \citet[p. 259]{ckt2}.
  \end{minipage}
  }
\end{figure}

Figure~\ref{fig:lf} shows industry-specific worker numbers from wartime to postwar.
With the decline in the total labor force during the war, the proportions of workers in commerce, domestic services, and other industries decreased.
Conversely, the mining, manufacturing, transportation, and communication industries saw an increase in their shares.
These trends indicate labor concentration in the military and aviation-related industries due to mobilization.
The number of workers in the agricultural and forestry sectors remained steady, implying that there was little demographic impact of labor mobilization in rural areas.

Despite the war ending, severe food shortages persisted until 1947 \citep{Shimizu1994}. 
Consequently, rural farmers preferred to continue farming on their traditional lands rather than leaving their villages to pursue opportunities in other industries. 
Although Fig.~\ref{fig:lf} indicates an increase in the labor force in the agricultural sector after the war, this increase was primarily due to the participation of returnees. Most of these returnees were individuals who transitioned to full-time farming, as opposed to their earlier status as part-time farm household members who earned their main income from jobs in other industries.
This shift to full-time farming was necessitated by the destruction of factories by air raids and the collapse of the urban labor market following the war's end \citep{msaf1946}.
Furthermore, while the demand for agriculture was driven by the insufficient food supply, limitations on available farmland made it difficult for people to relocate to rural areas and engage in farming.
Therefore, despite the statistical change in the labor force in the agricultural industry, the fundamental nature of farmers remained relatively unchanged throughout both wartime and the immediate postwar period. This suggests that population outflow from rural areas was a rare occurrence.

\subsubsection{Population change before and after air raids} \label{sec:bdr}

\begin{figure}[!t]

\begin{minipage}{0.5\textwidth}
\centering
\includegraphics[keepaspectratio, width=\textwidth]{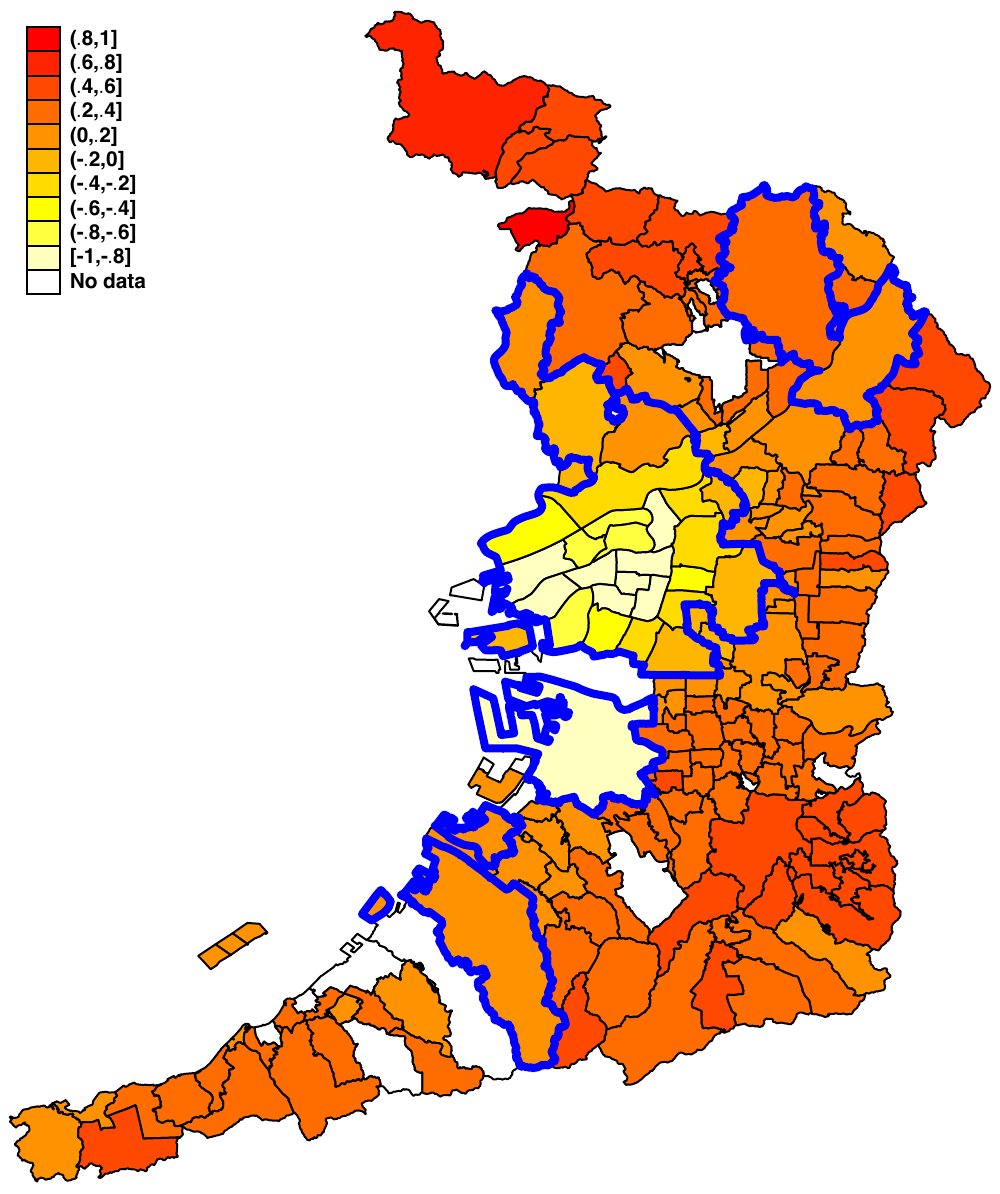}
\subcaption{1945}
\label{fig:osk1945}
\end{minipage}
\begin{minipage}{0.5\textwidth}
\centering
\includegraphics[keepaspectratio, width=\textwidth]{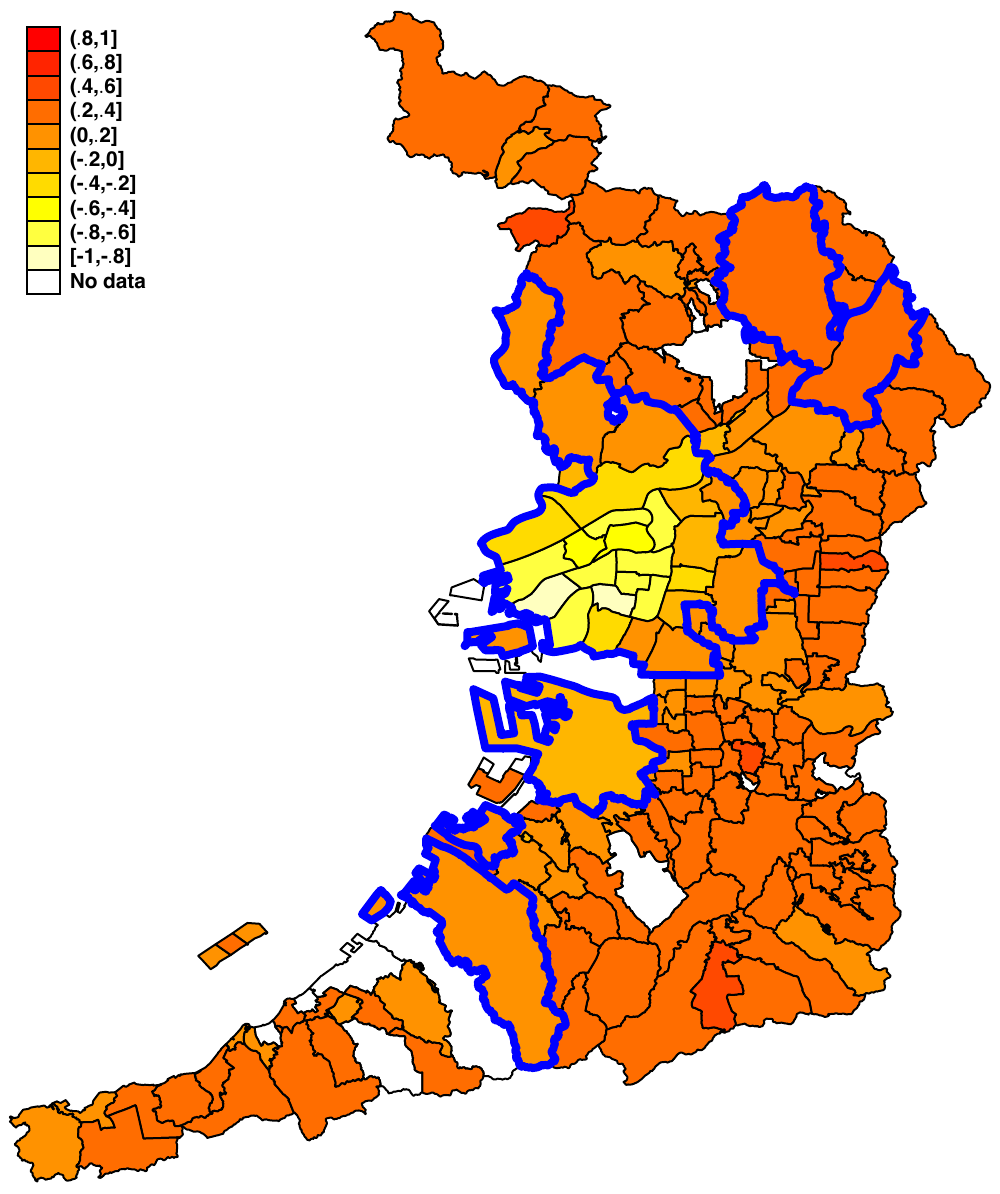}
\subcaption{1947}
\label{fig:osk1947}
\end{minipage}

\caption{Percentage change in population compared to 1944 in Osaka Prefecture}
\label{fig:osk}

\begin{minipage}{156mm}
\scriptsize{
Notes: 
The positive change rate indicates an increase relative to the population in February 1944, before the air raids intensified.
The data were based on surveys in November 1945 and October 1947.
Areas surrounded by thick lines denote cities.
The population is aggregated according to the 1950 municipal divisions.
Sources:
\citet[pp.46--59]{osaka1950} and the Ministry of Land, Infrastructure, Transport and Tourism (\url{https://nlftp.mlit.go.jp/ksj/index.html}, accessed on July 25, 2021).
}
\end{minipage}

\end{figure}

To summarize the population relocation, the limited number of people permitted to evacuate did not move to neighboring towns and villages but dispersed, relying on their relatives.
When the war ended, they gradually returned from their evacuated homes to their own cities.
Although there were many demobilized soldiers and returnees from overseas territories after the war, they did not congregate in one specific area.
Additionally, there was little rural-to-urban migration during both the wartime and immediate postwar periods.

Despite being a limited case study, historical data from Osaka, the only prefecture with municipality-level population data available before and after the air raids, corroborate these population changes.
Figure~\ref{fig:osk} shows the population changes in Osaka relative to 1944, the year before the air raids commenced.
In 1945, immediately after the air raids, the population dropped by up to 0.8\% in cities marked by thick lines on the map, while the population increased in all towns and villages.
This significant population increase in areas far from cities contradicts the notion of evacuation to areas closer to cities and the population outflow from rural areas.

In 1947, the urban population exhibited a recovery trend, indicating a temporary exodus from cities and the return of demobilized soldiers, repatriates, and evacuees.%
\footnote{For example, owing to intense air raids, Osaka City's population declined from 2,883,344 in 1944 to 1,102,959 in 1945.
The population recovered to 1,559,310 by 1947 \citep[pp. 46--59]{osaka1950}.}
Furthermore, the rate of population change was similar across towns and villages. Thus, the spatial distribution of the population reverted to the pre-air raid distribution, indicating no significant population concentration near cities when comparing the pre-war period with 1947.

\section{Empirical framework} \label{sec:emp}

\subsection{Potential mechanisms} \label{sec:mec}

We discuss three potential factors contributing to the far-reaching effects of the bombings on postwar fertility rates.%
\footnote{Rigorous identification of all mechanisms falls outside the main scope of this study.
Additionally, we focus on fertility changes during the postwar period rather than during the war.}
The most intuitive factor is the threat and corresponding fear caused by air raids.
This factor is a motivational factor for behavioral responses to violent incidents \citep{Williams2012}.
Even in towns and villages without direct damage, the sense of threat to oneself or one's family may have resulted in a decline in childbirth during the war.
Couples may have delayed childbearing due to challenges in emergency evacuating pregnant women and young children.
Furthermore, military conflicts can induce psychological stress that can lead to reduced reproductive behaviors \citep{Kidane1989}. 
In areas with neighboring cities experiencing air raids and severe damage, people might lose their desire for reproduction.
Psychological stress can adversely affect reproductive capacity \citep{Frankenberg2014, Nargund2015}.
If childbirth were postponed due to these mechanisms during wartime, fertility rates could rebound after the war. 
Hence, the threat of air raids could potentially have increased fertility in the postwar period.

Additionally, psychological theories support the idea that the fear of air raids in neighboring cities affects fertility.
Terror management theory suggests that a strong fear of violence can lead individuals to align with traditional values, like having more children and starting families \citep{Rodgers2005, Solomon2000}.
Attachment theory indicates that couples may seek greater proximity in response to threats such as disasters, reflecting attachment needs and potentially increasing fertility \citep{Cohan2002, Hazan1994}.
These behavioral changes due to psychological reactions can persist when individuals are exposed to intense fear \citep{Rodgers2005}.
Therefore, psychological theories suggest that whether the birth rate remains high or declines in the postwar period depends on the intensity of fear experienced during the air raids.

Another significant factor is the increase in socioeconomic insecurity attributed to air raids in neighboring cities.
However, the theoretical predictions regarding the effects of instability on fertility are inconsistent.
Increased instability can make childrearing more challenging, causing people to avoid having children.
The mechanism of delayed childbirth has also been interpreted from an economic perspective.
When parents expect their children to provide financial support in old age, childrearing is seen as an irreversible and long-term investment \citep{Aassve2021, Comolli2017}.
Insecurity makes such an investment riskier; therefore, couples are likely to postpone childbirth until after the war ends, potentially resulting in an increased birth rate during the postwar period.
Conversely, economic insecurity can enhance the value of economic support from children \citep{Cain1983}.
Before the mechanization of the agricultural sector, children were a crucial labor force in rural areas from a young age.
Thus, unless there is a clear increase in the risk of death, couples might choose to have more children during the war. Consequently, the fertility rate could decrease after the war.%
\footnote{
When child mortality increases and remains high for a long time, couples may have more children as a precaution to ensure that the desired number of children survives \citep{Frankenberg2014, Verwimp2005, Iqbal2010}.
This relationship between child mortality occurring outside the family and fertility decisions is known as the ``extrafamilial effect'' \citep{Nobles2015}.}
In summary, insecurity can have both positive and negative effects.

One specific potential factor is an increase in nationalism, which can ensure compliance with national population policies.
Starting in 1939, the Japanese government advocated population policies aimed at increasing the population, popularly known as ``\textit{umeyo fuyaseyo}'' (Beget and Multiply), to secure enough soldiers.
Moreover, given that air raids could boost national morale \citep{Jones2006, Jones2004}, those who directly witnessed bombings would likely have enhanced their morale compared to those who only learned about underreported casualties through newspapers and radio.
Therefore, inhabitants of towns and villages neighboring bombed cities might have actively followed the national policies, striving to have more children during the war due to rising nationalism.%
\footnote{In Japan during \wwii, patriotic behaviors were highly esteemed.
The special attack corps that sacrificed their lives by launching bodily attacks on their targets, known as \textit{kamikaze}, symbolized the values of that time.}
This increase in fertility would have ended with a defeat, limiting the postwar increase in fertility relative to other areas.

However, the mechanism involved in nationalism and patriotism was observed only in areas that did not suffer from intense fear.
This is because severe air raids shatter people's will and evoke avoidance responses to threats of harm and fear, rather than actions motivated by high national morale \citep{Gregor2000, Jones2006}.
This finding is consistent with the Japanese experience.
According to a survey report titled ``THE EFFECTS OF STRATEGIC BOMBING ON JAPANESE MORALE'' in \citet{USSBS}, air raids profoundly affected Japanese civilians’ hopes, fears, aspirations, sense of security, and attitude.%
\footnote{The United States Strategic Bombing Survey conducted a sample survey between November 10 and December 29 to assess the impact of the bombings on morale.
Using a two-stage sampling method to reduce sample selection bias, this survey involved 3,150 Japanese civilians aged 16–-70 years living in Honshu or western Kyushu (the main or southwestern islands of the Japanese archipelago).
The first stage of sampling covered 86\% of the population, and 76\% of those selected in the second stage eventually participated in the survey interviews.}
In this survey, 41\% of respondents reported that they were most worried about air raids during the war. 
For them, air raids were a more immediate threat than military defeats in the far-off Pacific Islands.
A Kyoto city official made the following statement: ``(in the respondent's area) in the last year of the war there was a marked changed in the emotional status of most people. 
Many became irritable, unstable, and critical of each other. . . . 
Word had gotten out that Kyoto would be wiped out by bombs and there was a near panic'' \citep[p. 35]{USSBS}. 
Additionally, the belief that Japan was winning the war because of false public announcements intensified people's sense of despair after the air raids.
Air raids accounted for 32\% of the factors that convinced rural inhabitants that Japan would lose the war \citep[p. 48]{USSBS}.
Conversely, 43\% felt better off after the war, primarily due to the cessation of air raids \citep[p. 22]{USSBS}.
These survey results corroborate the fact that the threat of air raids likely outweighed any increase in morale.

\subsection{Data} \label{sec:data}

For the empirical analysis, we compiled \WWII bombing damage and demographic data to create Kinki region town/village panel data for 1935 and 1947.
The Kinki region is ideal for studying bombing impacts owing to its cities' diverse damage.%
\footnote{It offers a higher variability of damages compared to other regions, making it more suitable for analysis.
For instance, the Tohoku region was less affected by air raids because of its distance from the bases of the Allied Forces.
In contrast, in the Kanto region, air raids against Tokyo City were devastating compared to other cities.}
Additionally, the Kinki region was spared from the atomic bombing, which could have different effects on fertility outcomes compared to conventional bombing.%
\footnote{The atomic bombings were targeted at Hiroshima City in the Chugoku region and Nagasaki City in the Kyushu region.
These cities experienced a high death toll, with estimated death rates of 23.2\% in Hiroshima and 8.8\% in Nagasaki among the 1944 population \citep[p. 283]{Nakamura1995}.}
We chose the years 1935 and 1947 as they represent the period just before and after \wwii, respectively, for which reliable demographic data are available.

To calculate the distances from each town or village to the cities, we used administrative divisions from 1950 because 1947 divisions were unavailable.%
\footnote{
Geographical information on the administrative divisions was obtained from the Ministry of Land, Infrastructure, Transport, and Tourism's download website (\url{https://nlftp.mlit.go.jp/ksj/index.html}, accessed on July 25, 2021).
The distances were computed based on the position of the center of gravity.
As Osaka, Kyoto, and Kobe have very large areas, we used the center of gravity of their closest districts when calculating the distances to these cities.
If towns and villages adjoined a city, the distance was considered zero.}
Hence, the variables were aggregated based on the 1950 divisions.
While some observations were omitted because of municipal mergers, our sample consists of 1,437 towns and villages, covering approximately 95\% of the 1,526 towns and villages in the Kinki region at that time.
This comprehensive dataset helps mitigate concerns related to sample selection bias.
Our sample did not include cities because we focused on the far-reaching indirect effects of bombings rather than their direct effects.

The main source of our data on bombing damage is \citet{esb1949}, who compiled human and building damages caused by air raids, categorized by damage scale in each city.
This source is considered the most reliable for air raids in Japan.
The data reported in \citet{USSBS} are widely used \citep[for example,][]{Davis2002}, but they only include casualties in the targeted cities of the Allied Forces.%
\footnote{However, the Economic Stabilization Board might have slightly overestimated the casualties because their purpose was to receive amnesty for wartime reparations by reporting wartime damage.
Therefore, we conducted additional estimates based on other data sources: the ``Final Reports of Records of the U.S. Strategic Bombing Survey'' in \citet{USSBS} and the ``List of Casualties from Air Raids on War Damaged Cities in Japan'' in \citet[pp. 71--72]{sensai1962}.
The estimation results remain unchanged, supporting the validity of the main results.
See Online Appendix B for further details and other robustness checks.}
We adopted the number of deaths from bombings as a measure of the intensity of the war damage.
Using the distances to bombed cities and their death tolls, we calculated the total number of deaths due to bombings at various distances from each town or village.%
\footnote{Focusing on carpet bombings on civilians, we exclude damages in cities where only precision bombings on military facilities and plants were carried out, even if there were unfortunate deaths.
Consequently, the number of deaths in Suzuka, Kainan, and Nara was regarded as zero.
However, our estimation results remain unchanged if we include the damage in these cities (not reported).}
This is our key independent variable for capturing the far-reaching effects of bombings on postwar fertility.

For demographic data at the municipality level in 1935 and 1947, we obtained male and female populations from the \textit{Population Census of Japan} \citep{census1935pp, census1947pp}, and the number of live births from the \textit{Vital Statistics of Japan by Municipalities} \citep{mvs1935, vsej1947}.
These data sources were based on population census data.%
\footnote{Despite the chaos of war, the government accurately captured the population because the registration of the resident register was required to receive rationed foods and goods, which were essential for living at that time \citep[p. 89]{rra1950}.}
Our outcome measure of fertility level was the crude birth rate, defined as the number of live births per 1,000 people.
We also used the log of the population and the male-to-female sex ratio as the main control variables.
Table~\ref{tab:sum} presents summary statistics for these variables.

\begin{table}[t]
\tabcolsep =2mm
\begin{center}
\caption{Summary statistics}
\label{tab:sum}
\footnotesize

\begin{tabular*}{160mm}{l@{\extracolsep{\fill}}D{.}{.}{1}D{.}{.}{3}D{.}{.}{1}D{.}{.}{1}D{.}{.}{3}}
\toprule

\hspace{40mm} &\multicolumn{1}{c}{Mean}&\multicolumn{1}{c}{Std. Dev.}&\multicolumn{1}{c}{Min}&\multicolumn{1}{c}{Max}&\multicolumn{1}{c}{Observations}\\

\hline

Crude birth rate
&31.80&4.89&10.12&56.63&2,874\\

\textit{Sex ratio}
&0.97&0.08&0.56&2.00&2,874\\

ln\textit{Population}
&8.03&0.58&5.57&10.28&2,874\\

\textit{Proximity to city}
&0.59&0.49&0&1&2,874\\

\textit{City population}
&7.39&6.27&0.00&15.98&2,874\\

\\
\multicolumn{6}{l}{Death toll from bombing in nearby cities (only non-zero values)}\\

\hspace{9pt}$\leq$ \hspace{5pt}5 km
&1,413.35 &2,797.71 &1&10,663 &207 \\
\hspace{9pt}$\leq$ 10 km
&2,155.59 &3,603.63 &1&18,172 &355 \\
\hspace{9pt}$\leq$ 15 km
&2,499.21 &3,996.18 &1&18,432 &577 \\
\hspace{9pt}$\leq$ 20 km
&2,707.69 &4,208.79 &1&19,888 &808 \\
\hspace{9pt}$\leq$ 25 km
&3,316.16 &4,682.16 &1&19,888 &974 \\

\bottomrule
\end{tabular*}

{\scriptsize
\begin{minipage}{156mm}
Notes: Crude birth rate is defined as the number of live births per 1,000 population.
The sex ratio is calculated as the number of men divided by the number of women.
Proximity to the city is the dummy variable, taking one if there was at least one city within 15 km.
The city population is defined as the inverse hyperbolic sine of the total population of cities within 15 km.
Sources: \citet{esb1949, census1935pp, census1947pp, mvs1935, vsej1947}.
\end{minipage}
}
\end{center}
\end{table}

\subsection{Baseline identification strategy} \label{sec:model}

This case study of Japan during \WWII offers the following advantageous settings:
First, bombing targets and damages were determined independently of any characteristics of the surrounding towns and villages, even if they correlated with some determinants of the fertility rate in bombed cities.
Moreover, many factors created variance in the magnitude of damage to cities \citep{Davis2002}.
Thus, this exogenous treatment enables us to estimate the far-reaching effects of bombings on the postwar crude birth rate in towns and villages without being disturbed by other confounding factors, such as the number of demobilized soldiers.
The exogeneity also relaxes the common trend assumption in our analysis.

Second, the geographical location of mainland Japan as an island nation in the Far East inherently limited the Allied Forces' modes of aggression almost exclusively to aerial bombardment.%
\footnote{
The Allied Forces landed only in Okinawa Prefecture during \wwii.
Naval bombardment hit several cities on the mainland but caused negligible damage, accounting for only 0.5\% of the total bombing damage \citep[pp. 278, 284]{Nakamura1995}.
Considering that only Shingu City suffered such attacks in the Kinki region, our analysis focused on air-raided damage.
}
In continental nations or regions, adversarial forces have the option of invading a target city through other land territories, which may directly affect fertility rates in areas along the path.
Additionally, when a nation is subjected to a multitude of assault methods, the effects on fertility can vary depending on the attack type, even if the measured damage sizes are identical.
Japan's wartime damage, limited to air raids, avoids these possibilities and ensures the validity of our empirical analysis.
The relatively short period of air raids against the entire city, spanning only from early 1945 to the end of the war, is also beneficial for our analysis because the influence of war damages on decision-making regarding childbirth might change as the war drags on \citep{Iqbal2010}.

Finally, the information control exercised by the Japanese government during wartime prevented the spread of indirect mental shocks due to air raids nationwide, which helps us identify the far-reaching effects.
With the intention of maintaining national morale, official narratives markedly overstated military success and understated domestic casualties.
Consequently, populations residing at a considerable distance from bombed cities remained largely uninformed of air raids and thus were not influenced by them.
In contrast, those close to bombed cities could see and hear the firsthand destruction caused by aerial bombardment.
Therefore, bombings were likely to have more potent indirect effects on areas closer in proximity but did not affect areas that were too far away.
This relationship between proximity and the magnitude of the effects enables us to distinguish the localized impacts of air raids from national-level macro shocks.

To investigate the far-reaching effects of war damage on fertility, we employ a fixed-effects model in the spirit of the DID approach.
Our baseline model is as follows:
\begin{eqnarray}
\textit{CBR}_{it} = \alpha + \beta (\textit{Bombing}_{i}\times\textit{Postwar}_{t}) + \gamma\textit{Postwar}_{t} + \vx'_{it} \vdelta + \nu_{i} + \varepsilon_{it} \label{eq:base}
\end{eqnarray}
In the equation above, $i$ represents towns and villages numbered from 1 to 1,437, and $t$ represents the years 1935 and 1947.
The dependent variable $\textit{CBR}_{it}$ stands for the crude birth rate.
$\textit{Bombing}_{i}$ indicates the intensity of air raids in neighboring areas during \wwii, measured by the number of deaths (in thousands) from bombings in cities within 15 km.
$\textit{Postwar}_{t}$ is a dummy variable that equals one if the year is 1947, representing the year-fixed effect.
The interaction term $\textit{Bombing}_{i}\times\textit{Postwar}_{t}$ captures the far-reaching effects of bombing on postwar fertility.
$\vx'_{it}$ is a vector of control variables, $\nu_{i}$ represents town and village fixed effects, and $\varepsilon_{it}$ is a random error term.
The year-fixed effect controls for all year-specific unobservable confounders, allowing us to identify the indirect effects of the bombings without being influenced by national level macro shocks, including those related to the war.
Additionally, town and village fixed effects account for heterogeneity across municipalities.

A potential concern in our empirical context is that many individuals who fled from bombed cities may have sought refuge in neighboring villages and established their homes there.
If a direct experience of air raids altered their desired number of children, our key variable could potentially encompass a combination of both direct and indirect effects of bombings.
Additionally, returnees might have chosen to settle in proximity to cities.
These concerns contrast with historical evidence, which suggests that people evacuated or returned to the countryside where their relatives resided rather than to nearby rural areas.
Nonetheless, we address this possibility by incorporating the male-to-female sex ratio and the logarithm of the population as control variables in our estimation model.
These variables help account for the impact of population relocation and demographic characteristics on our analysis.%
\footnote{For additional insights, \citet{OgasawaraKomura2021} has emphasized the importance of considering the role of sex ratio in assessing the effects of demobilization on fertility rates.}

Another issue to consider is the geographical proximity to a city.
The definition of the bombing damage variable inherently assigns a value of zero to towns and villages far from any city.
In contrast, for observations in close proximity to cities, this variable takes a positive value if the nearby city is subjected to bombing and zero otherwise.
Therefore, our estimates could be biased if the impact of geographical proximity on birth rates varied after the war.%
\footnote{
For instance, male workers in rural areas near cities might have had higher mobility.
The decline in the male labor force could have placed an additional burden on female agricultural workers, leaving them with little capacity to bear and raise children.
The challenge women faced in balancing agricultural responsibilities with childbearing and child-rearing during that period is supported by the research finding of \citet{Okado1996}, which indicates that infant mortality rates increased in households with larger expanses of arable land.
}
To address this potential confounding factor, we introduce an interaction term into our estimation model as a control variable. 
This term combines a dummy variable, set to one if a city is located within a 15-kilometer radius (\textit{Proximity to city}), with the postwar dummy variable.
Additionally, larger cities might have had a more significant influence on neighboring towns and villages.
To account for this concern, we adopt an estimation equation that controls for the total population of cities within 15 km, replacing the proximity dummy.%
\footnote{When taking the natural logarithm of the city population, we apply the inverse hyperbolic sine transformation (\ie, ln$(x+\sqrt{x^2+1})$ for a random variable $x$) to retain observations whose value is zero.}
These adjustments ensure that we consider the potential influence of geographical proximity on our analysis, particularly with regard to postwar birth rates.

\section{Estimation results} \label{sec:res}

\subsection{Baseline results} \label{sec:base}

Table~\ref{tab:result} presents these results of our analysis.
In column (1), we observe the regression of the crude birth rate on the key independent variable, bombing damage.
In column (2), we introduce the sex ratio as an additional control variable, and in column (3), we include the log of the population.
Our baseline specification, found in column (4), encompasses the sex ratio, log of population, and a dummy variable representing proximity to a city. This specification maintains town and village fixed effects and postwar dummy variables.
Column (5) displays the results when we incorporate the population of nearby cities instead of the proximity dummy.
Once again, town and village fixed effects and postwar dummy variables are included in all specifications.

In column (1), we observe that the estimated coefficient of the interaction term between bombing damage and the postwar dummy variable is not only positive but statistically significant at the one percent level.
This result indicates that air raids in neighboring cities during \WWII had a discernible impact on the crude birth rate in towns and villages during the postwar period.
This aligns with our hypothesis that war damage indirectly influences fertility in areas not directly affected but near damaged regions.
Columns (2) and (3) demonstrate that the estimation results remain consistent even when we control for demographic variables such as the sex ratio and population.
These findings confirm the reliability of our results.

Column (4) maintains a significantly positive estimated coefficient for the key independent variable, while the proximity dummy's coefficient is positive but statistically insignificant.
These results suggest that the relationships between bombing damage and crude birth rates are not due to geographical proximity to a city but rather stem from the far-reaching effects of air raids.
Our baseline results indicate that for every 1,000 additional deaths caused by air raids within a 15-kilometer radius, the crude birth rate increased by 0.245 permil in 1947.
Replacing the proximity dummy with the nearby cities' total population in column (5) yields consistent estimation results.
Hence, the size of nearby cities does not seem crucial for pre- and post-WWII fertility rates.

\begin{table}[t]
\tabcolsep = 2mm
\begin{center}
\caption{Effects of bombing damage within 15 km on fertility}
\label{tab:result}
\footnotesize

\begin{tabular*}{160mm}{l@{\extracolsep{\fill}}ccccc}
\toprule

&(1)&(2)&(3)&(4)&(5)\\

\hline

\textit{Bombing}$\times$\textit{Postwar}
&0.234***&0.234***&0.268***&0.245***&0.218***\\
&(0.047)&(0.047)&(0.054)&(0.055)&(0.099)\\

\textit{Sex ratio}
&&1.177&2.091&1.956&2.0030\\
&&(2.770)&(2.844)&(2.848)&(2.839)\\

ln\textit{Population}
&&&$-$1.965&$-$2.276&$-$2.269\\
&&&(1.497)&(1.531)&(1.531)\\

\textit{Proximity to city}$\times$\textit{Postwar}
&&&&0.541&\\
&&&&(0.359)&\\

\textit{City population}
&&&&&$-$0.280\\
&&&&&(1.137)\\

\textit{City population}$\times$\textit{Postwar}
&&&&&0.045\\
&&&&&(0.033)\\

\hline

Town and village fixed effects
&Yes&Yes&Yes&Yes&Yes\\

Postwar fixed effect
&Yes&Yes&Yes&Yes&Yes\\

Observations
&2,874&2,874&2,874&2,874&2,874\\

\bottomrule
\end{tabular*}

{\scriptsize
\begin{minipage}{156mm}
Notes: The dependent variable is the crude birth rate (permil).
The independent variable \textit{Bombing} is the number of deaths (in thousands) due to the bombing in cities within 15 km.
The \textit{Sex ratio} is the male-to-female sex ratio, the ln\textit{Population} is the log of population, the \textit{Proximity to city} is the dummy variable taking one if there were cities within 15 km, and the \textit{City population} is the inverse hyperbolic sine of the total population of cities within 15 km.
***, **, and * represent statistical significance at the 1\%, 5\%, and 10\% levels, respectively.
Standard errors clustered at the town--village level are in parentheses.
\end{minipage}
}
\end{center}
\end{table}

\subsection{Geographical extent of effects} \label{sec:ext}

\afterpage{
\begin{figure}[t]

\begin{minipage}{0.5\textwidth}
\centering
\includegraphics[keepaspectratio, width=\textwidth]{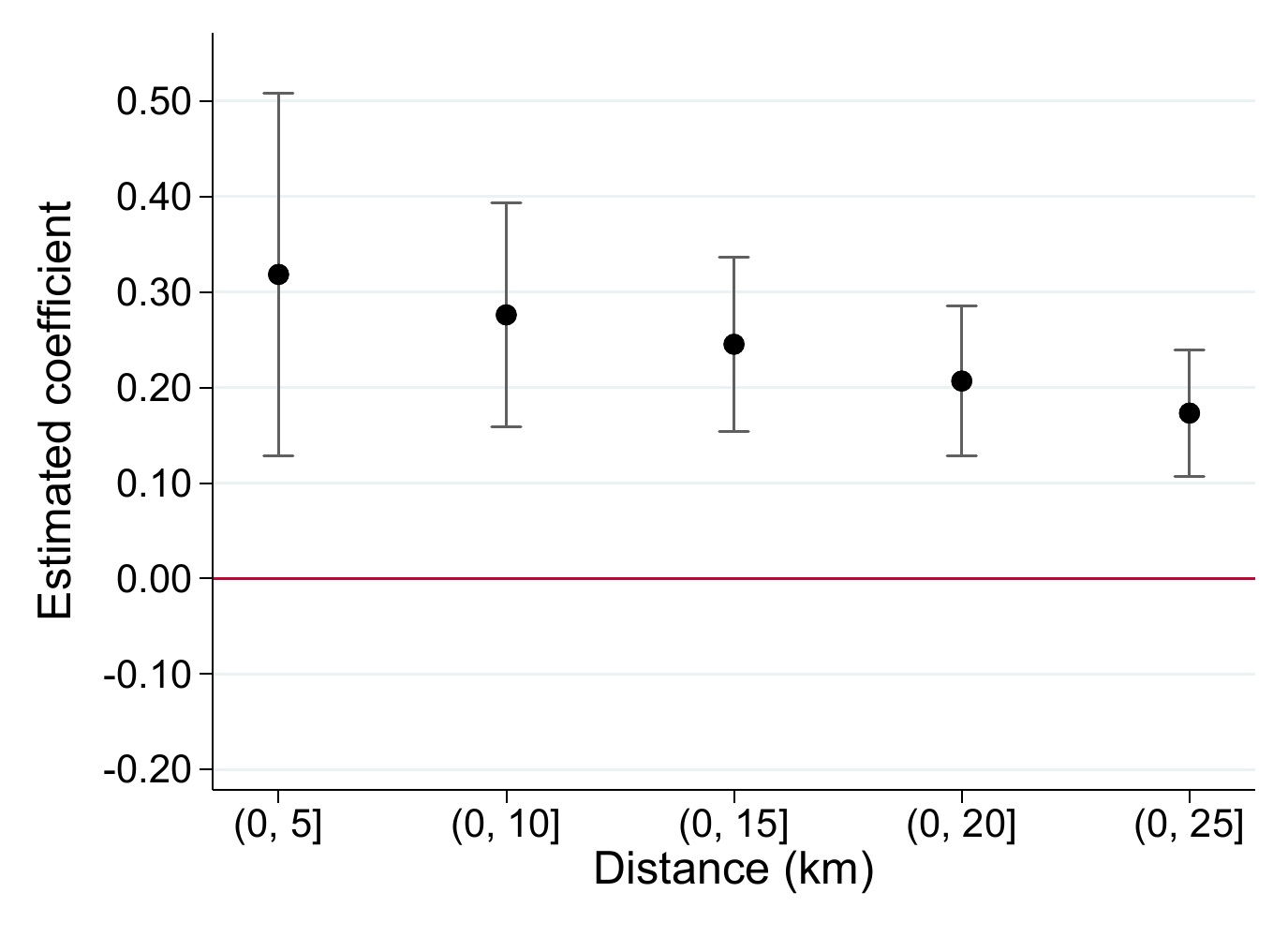}
\subcaption{Continuous distance}
\label{fig:res_main}
\end{minipage}
\begin{minipage}{0.5\textwidth}
\centering
\includegraphics[keepaspectratio, width=\textwidth]{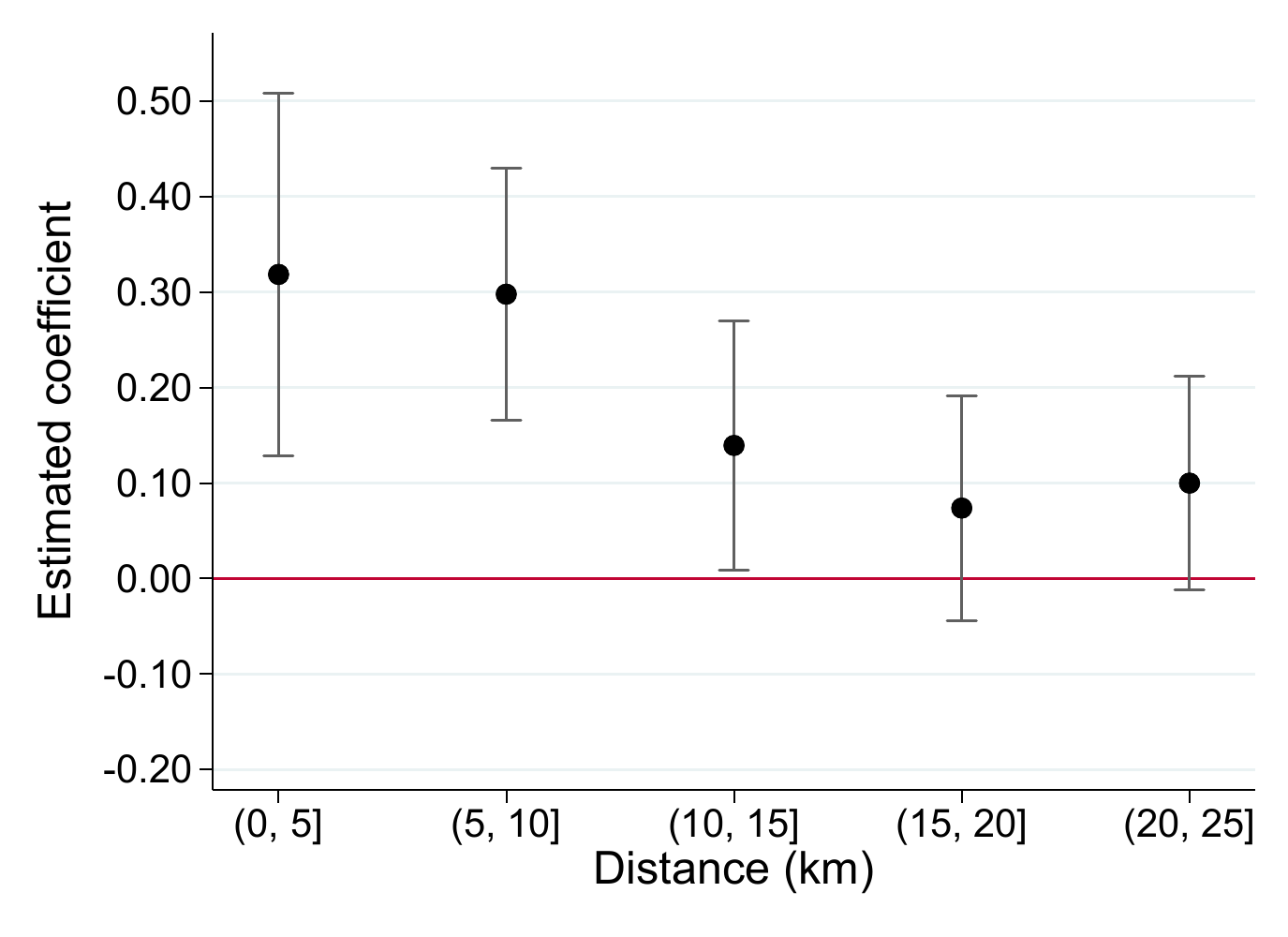}
\subcaption{Graded distance}
\label{fig:res_sub}
\end{minipage}

\caption{Effects of bombing damage within various distances on crude birth rate}
\label{fig:res}

\begin{minipage}{156mm}
\scriptsize{
Notes: 
The dotted and solid lines indicate the point estimates and their 90\% confidence intervals, respectively.
The dependent variable is the crude birth rate (permil).
The key independent variable is the death count (in thousands) due to bombing in cities within various distances, interacting with the postwar dummy variable.
Figure~\ref{fig:res_sub} controls for the bombing damage at shorter distances.
The control variables are the male-to-female sex ratio, log of population, and dummy variable, taking one if there were cities within each distance.
The town and village fixed effects and postwar dummy are also included.
A robust standard error is clustered at the town--village level.
}
\end{minipage}

\end{figure}
}

To explore the geographical reach of fertility effects, we assessed bombing damage impact at different distances.
The estimating equation used here is consistent with the baseline specification used in column (4) of Table~\ref{tab:result}, but it employs varying distances to capture the effects of bombing instead of the fixed 15-kilometer radius.
Figure~\ref{fig:res_main} illustrates the results, with the dotted line representing the estimated coefficient of the total death toll (in thousands) from air raids in cities within 5, 10, 15, 20, and 25 km, all of which were positive.
Furthermore, the solid lines depicting the 90\% confidence intervals indicate that the results hold statistical significance at all distances.
These findings are consistent with the baseline results, affirming the presence of far-reaching effects of bombing on fertility.
The declining coefficients imply that air raids had a more pronounced impact on fertility in towns and villages near bombed cities.

We also estimated the effects of bombings in cities within 0--5, 5--10, 10--15, 15--20, and 20--25 km to explore the detailed relationship between distance to air raids and their impact.
However, the impact at a particular distance might have varied based on the severity of bombing damage in closer cities. For instance, the effect of bombing within 5--10 km may have been smaller in towns and villages that experienced severe air raids within five km. 
To address this, we controlled for bombing damage within a shorter distance.
Thus, when we estimate the effect of bombings within 10--15 km, the estimation equation includes bombing damages within 10 km.
The results are presented in Figure~\ref{fig:res_sub}.
The estimated coefficient of bombing damage remained significantly positive at distances of 5--10 km and 10--15 km, while those at 15--20 km and 20--25 km were statistically insignificant.
These findings imply air raids influenced the birth rate within 15 km but showed no significant effects beyond.

The estimation results are shown in Figs.~\ref{fig:res_main} and \ref{fig:res_sub} are complementary.
Although Fig.~\ref{fig:res_main} shows significant results at all distances, it does not provide specific insight into whether air raids within the extended distances had significant effects.
For instance, the estimated effect within 25 km, although significant, may have been influenced by bombings within 5 km.
The analysis results presented in Fig.~\ref{fig:res_sub} address this issue.
In contrast, using graded distances prevents us from estimating comprehensive effects.
Thus, Fig.~\ref{fig:res_sub} does not encompass the effects of air raids within distances shorter than the specified range.
To counterbalance this limitation, we rely on estimates in Fig.~\ref{fig:res_main}.
Therefore, it is essential to interpret both sets of results in tandem. These combined findings indicate that air raids within 15 km significantly influenced the postwar crude birth rate.

\subsection{Heterogeneity in effects} \label{sec:hetero}

We also investigated the heterogeneity of the far-reaching effects of air raids.
While our baseline model assumes linearity, the effects of bombings may vary depending on damage intensity.
Thus, we introduced a categorical variable representing clusters of damage intensity to capture these heterogeneous effects. 
The estimation equation is as follows:
\begin{eqnarray}
\textit{CBR}_{it} = \alpha + \sum_{k=1}^{5}\beta_{k}(C_{ki} \times\textit{Postwar}_{t}) + \gamma\textit{Postwar}_{t} + \vx'_{it} \vdelta + \nu_{i} + \varepsilon_{it} \label{eq:hetero}
\end{eqnarray}
where $C_{ki}$ is an indicator variable that takes on a value of one if $\textit{Bombing}_{i}$, which represents the number of deaths (in thousands) due to bombings in cities within 15 km of $i$ is classified into the $k$-th cluster of the damage intensity.
The first to fifth clusters consist of death tolls of 1--26, 81--269, 416--881, 1,271--1,895, and 6,789--18,432, respectively.%
\footnote{These clusters are generated by the k-means clustering algorithm, with quintile groups as the initial clusters for repeatability.}
The estimated parameter $\hat{\beta_{k}}$ is interpreted as the mean effect of bombings on the crude birth rate in the $k$-th cluster, reflecting the heterogeneity of the effects depending on the damage intensity.

\begin{figure}[!t]

\centering
\includegraphics[keepaspectratio, width=0.75\textwidth]{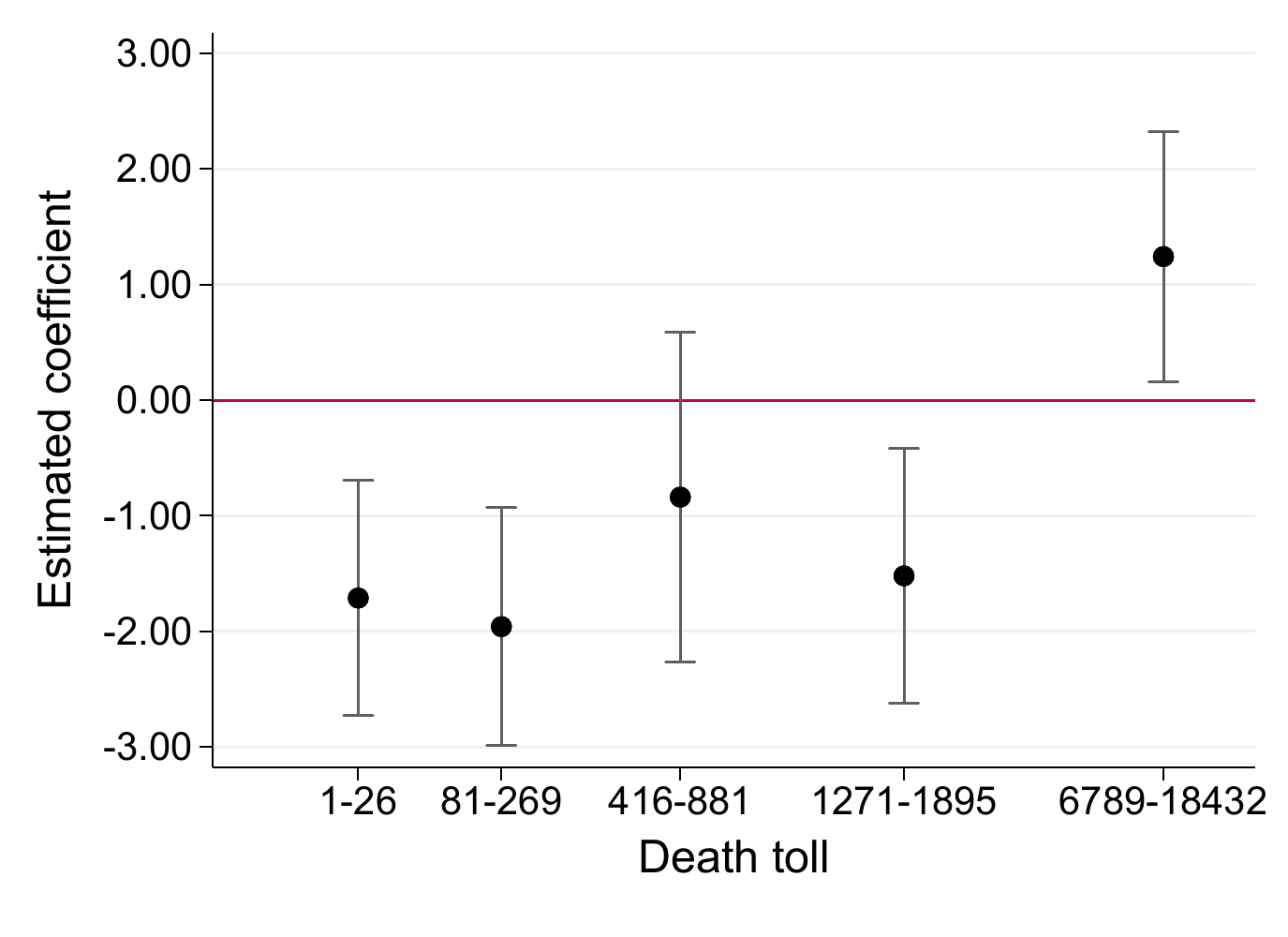}
\caption{Effects of bombing on fertility by damage size within 15 km}
\label{fig:res_rob}

\begin{minipage}{156mm}
\scriptsize{
Notes: 
The dotted and solid lines indicate the point estimates and their 90\% confidence intervals, respectively.
The dependent variable is the crude birth rate (permil).
The key independent variables are the dummy variables representing each cluster of bombing damage intensity within 15 km (1--26, 81--269, 416--881, 1,271--1,895, and 6,789--18,432 deaths), interacting with the postwar dummy variable.
The numbers of observations in each cluster are 320, 232, 130, 224, and 248, respectively.
The control variables are the male-to-female sex ratio, the log of population, and a dummy variable representing proximity to a city.
The town and village fixed effects and postwar dummy are also included.
A robust standard error is clustered at the town--village level.
}
\end{minipage}

\end{figure}

Figure~\ref{fig:res_rob} shows the results estimated using Equation~(\ref{eq:hetero}), in which the control variables include the sex ratio, population, and proximity dummy.
The significant positive effect at the strongest intensity corresponds to our baseline results.
However, the effects in the other clusters were negative.
This result difference suggests that the bombing's far-reaching effects on the crude birth rate varied with the scale of damage.
Small-scale air raids may have negatively impacted postwar fertility, while large-scale air raids may have had a positive influence.

The different effects imply distinct channels by which bombing indirectly impacted nearby postwar fertility.
While the role of insecurity is ambiguous because of its potential positive and negative effects, these results align with the mechanisms of threat and nationalism.
Relatively small-scale air raids could have bolstered nationalism but might not have posed a major threat.
Hence, wartime population policies promoting childbirth may have led to a postwar fertility rate decrease.
In contrast, large-scale air raids instilled fear in people and destroyed national morale, leading to a positive impact on postwar fertility.

The mean effect for the first to fourth clusters was $-1.507$, roughly equivalent to 4.74\% of the sample mean crude birth rate.
While this is interpreted as a negative, far-reaching effect of the bombing, the treatment group is inevitably affected by proximity to a city.
Given that the estimated coefficient of the proximity dummy variable is 1.603, the combined effects amount to 0.096.
This calculation suggests that the adverse, far-reaching effect on postwar fertility offsets the positive impact of unobservable factors associated with proximity.
For the cluster with the most severe damage, the total effect was calculated as 2.844, which is approximately 8.94\% of the mean crude birth rate in our sample.
The magnitude of this effect is over 10 times larger than the estimated coefficient of 0.275 for the postwar dummy, indicating the significant roles of air raids and city proximity in shaping postwar fertility, even in undamaged areas.

\subsection{Quasi-experimental approach} \label{sec:dir}

To clarify distinctions in far-reaching effects through potential channels, we use a quasi-experimental approach based on Allied Forces flight paths.
They flew from military bases in the Mariana Islands, southeast of Japan, when bombing Japanese cities.%
\footnote{A typical flight path of the Allied Forces is displayed in Online Appendix A.}
Hence, those in the southeast likely felt heightened fear due to visible bombers and frequent air-raid sirens.
However, other potential mechanisms, like increased insecurity and shifts in national morale, were unrelated to neighboring bombed city directions.
Given the exogenous factors of Allied Forces’ departure direction and town/village locations, we can assess bombing's far-reaching fertility effects through the lens of threat and fear, using the direction from bombed cities.

The analytical model that incorporates the direction is presented as follows: 
\begin{eqnarray}
\textit{CBR}_{it} = \alpha + \beta (\textit{Bombing}_{i} \times \textit{Postwar}_{t} \times I_{di}) + \gamma\textit{Postwar}_{t} + \vx'_{it} \vdelta + \nu_{i} + \varepsilon_{it} \label{eq:cir}
\end{eqnarray}
Here, we use $d = 1, ..., 360$ to represent the direction angle.
The indicator variable $I_{di}$ takes on the value of one if a town or village $i$ is located within a range of $d \pm 22.5$ degrees of the nearest bombed city.
The link between bombing damage and direction is not strictly deterministic because direction is determined solely based on the location of the nearest bombed city, whereas damage accounts for the total damage in all cities within 15 km.
Nevertheless, our estimates effectively capture the overall trend of far-reaching effects based on direction.
The control variables and fixed effects remain consistent with those in the baseline specifications.

\begin{figure}[!t]

\centering
\includegraphics[keepaspectratio, width=0.75\textwidth]{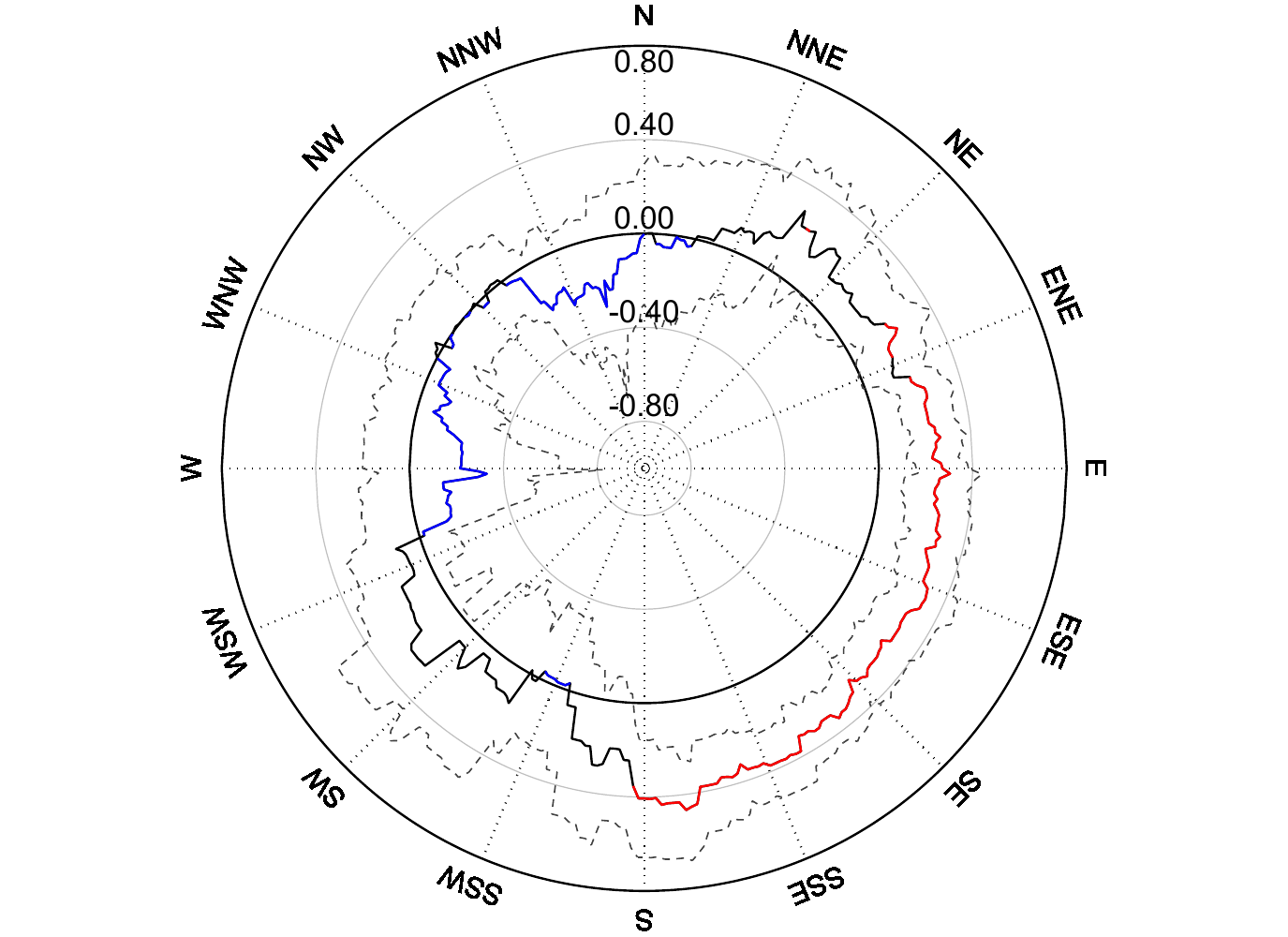}
\caption{Effects of bombing on fertility by direction from the nearest city}
\label{fig:res_circle}

\begin{minipage}{156mm}
\scriptsize{
Notes: 
The solid and dashed lines indicate the point estimates and their 90\% confidence intervals, respectively.
Two villages surrounded by cities or the sea in all directions are excluded.
The dependent variable is the crude birth rate (permil).
The key independent variable \textit{Bombing} is the number of deaths (in thousands) due to the bombing in cities within 15 km, which is interacted by the postwar dummy and indicator variables representing the angle of direction from the nearest city (including $\pm22.5$ degrees).
The control variables are the male-to-female sex ratio, the log of population, and a dummy variable representing proximity to a city.
The town and village fixed effects and postwar dummy are also included.
A robust standard error is clustered at the town--village level.
}
\end{minipage}

\end{figure}

Figure~\ref{fig:res_circle} shows a visualization of the results in all directions.
The analysis revealed three distinct groupings of the estimated effects on fertility based on geographical orientation.
First, we observed positive and statistically significant effects between the South and East.
Second, positive but statistically insignificant relationships were found between the East and North, as well as between the West and South.
Finally, negative impacts were identified between the North and West.

These findings align with our hypothesis, derived from potential mechanisms and varying damage intensity interpretations.
The significantly positive effects on the southeastern side, directly under the flight paths, support the idea that air raids induced significant fear, leading to increased postwar fertility.
Conversely, the effects were negative on the northwestern side.
In this direction, air raid-related fear was likely weakest, as it faced away from the southeast.
Thus, without intense fear, bombings may have far-reaching effects, reducing postwar fertility rates.
The statistically insignificant positive effects in other directions may result from offsetting positive and negative impacts.

Further analysis examined far-reaching effects in the southeast and other orientations based on damage intensity.
The estimation model is as follows: 
\begin{eqnarray}
\textit{CBR}_{it} = \alpha + \sum_{k=1}^{3}\beta_{sk}(C_{ki} \times\textit{Postwar}_{t}\times I_{si}) + \gamma\textit{Postwar}_{t} + \vx'_{it} \vdelta + \nu_{i} + \varepsilon_{it} \label{eq:hetero}
\end{eqnarray}
where $s$ denotes a directional category, indicating either southeast or otherwise.
The categorical variable $I_{si}$ specifies whether a town or village $i$ is situated on the southeastern side of a city affected by the air raids. 
Specifically, $i$ is classified as being in the southeast if it is positioned between the south and east of at least one of the bombed cities within 15 km.
An observation $i$ that does not meet this criterion is defined as belonging to a different directional category.
Under the assumption that only the fear of bombing differed between these directions, the difference in the estimated coefficients is interpreted as the far-reaching effect of fear.
To ensure a sufficient number of observations in each cluster, we grouped the damage intensity into three clusters (1--583, 855--1895, and 6,789--18,432 death tolls).

\begin{figure}[!t]

\centering
\includegraphics[keepaspectratio, width=0.75\textwidth]{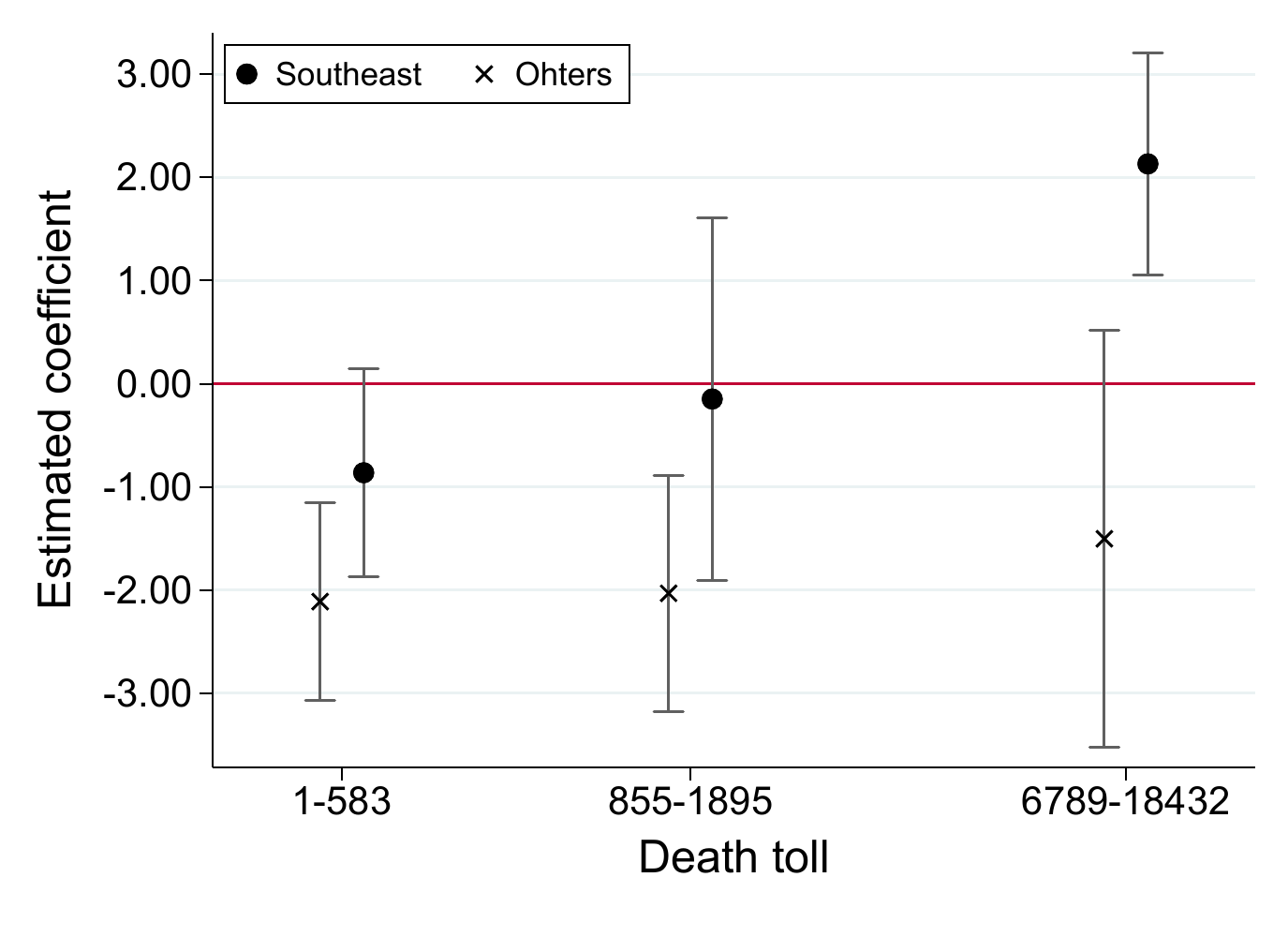}
\caption{Effects of bombing on fertility by damage size and direction}
\label{fig:res_robdir}

\begin{minipage}{156mm}
\scriptsize{
Notes: 
The dotted and solid lines indicate the point estimates and their 90\% confidence intervals, respectively.
The dependent variable is the crude birth rate (permil).
The key independent variables are the dummy variables representing each cluster of bombing damage intensity within 15 km (1--583, 855--1895, and 6,789--18,432 deaths), interacting with the postwar and southeast dummy variables. 
The numbers of observations in each cluster are 133, 31, and 95 in the southeast and 196, 93, and 29 in others, respectively.
The control variables are the male-to-female sex ratio, the log of population, and a dummy variable representing proximity to a city.
The town and village fixed effects and postwar dummy are also included.
A robust standard error is clustered at the town--village level.
}
\end{minipage}

\end{figure}

Figure~\ref{fig:res_robdir} presents the results.
For towns and villages situated southeast of the bombed cities, the estimated coefficient is negative in the cluster with the least impact.
However, a noticeable upward trend is observed, culminating in a significant positive effect in the most intense clusters.
While it is impossible to exclude insecurity as a contributing factor to this upward trend, these results suggest that severe air raids amplified the sense of threat, leading to more childbirths after the war.

Conversely, non-southeastern areas consistently yield negative outcomes in all damage clusters. This pattern implies the dominant role of insecurity or nationalism in shaping the far-reaching effects of the bombing. 
Furthermore, larger coefficients in severe damage clusters suggest fear as a significant factor in the effect, even outside the southeast.
Even in the cluster with the highest damage level, the increase in effect size was minor.
Hence, fear intensity and its impact on postwar fertility depend on direct sensory air raid experiences. Alternatively, the slight non-southeastern increase may result from the insecurity caused by bombings.

Although the estimated coefficients represent the cumulative effect through various mechanisms, the difference between southeast and other directions' coefficients highlights the pure fear effect.
The differences in effects across the damage intensity clusters exhibited a gradual increase, with values of 1.25, 1.88, and 3.63, respectively, from the least to the most intense clusters. 
This widening gap supports the high fear susceptibility in southeastern city areas and its far-reaching positive effects on postwar fertility.

\section{Conclusion}\label{sec:con}

This study unveils significant postwar fertility impacts from \WWII air raids in Japan's Kinki region.
Using panel data of 1935 and 1947 town- and village-level fertility rates and air raid death tolls, we studied impacts in areas near bombed cities, unaffected by bombings.

Through regression analysis grounded in robust historical evidence, we have demonstrated that the bombing’s impact extended to a radius of 15 km.
Intense air raids could have positively influenced postwar fertility, whereas less severe air raids had the opposite effect.
Our simple calculations suggest that both effects correspond to approximately 5\% of the mean crude birth rate at that time.
However, our focus was limited to the 1947 effects, and aerial bombardment may have had more pronounced impacts at different times.

Furthermore, our quasi-experimental approach suggests a fertility increase in municipalities southeast of bombed cities, where Allied bombers likely passed during raids.
While revealing the complex mechanisms linking air raids to indirect fertility impacts is challenging, this finding suggests that wartime threats and fears may have contributed to a postwar birth rate increase.

The far-reaching effects in this study may differ from those of \citet{Harada2022}, who studied air raids in Tokyo and linked community-level social capital development to direct damage.
However, indirect impacts can vary based on measures and timeframes.
Hence, this study offers a new perspective on fertility impacts, underscoring the need for broader research on air raid repercussions.

\section*{Acknowledgment}\label{sec:ack}
We would like to extend our heartfelt appreciation to the participants of the Economics Workshop at Meiji University, CIGS Meeting, CSG Workshop at Meiji University, and the Economics and Game Theory Seminar at Tokyo University of Science for their valuable insights and feedback.
This study received support from a Grant-in-Aid for Early-Career Scientists [Grant Number: 22K13440] and a Research Fellowship for Young Scientists [20J21183].
All errors or omissions in this work are solely our responsibility.

\bibliographystyle{./bib/draft}
\bibliography{./bib/ref}

\appendix

\clearpage



\section*{Supplementary material}

# Appendices

# Appendix A  Allied flight path and responses to air rids

Figure A.1a shows the location of the Kinki region, and Fig. A.1b shows an example of flight paths taken by Allied bombers when they attacked Kinki. Flying from the Mariana islands, including Guam, Saipan, and Tinian, located southeast of Japan without long detours, passed through the southeastern municipalities of the cities. In addition, they could not fly in other directions because they were only able to depart from military bases in the Mariana Islands to conduct air raids. As flight paths depend on exogenous factors, such as the aeronautical capabilities of the bombers and geographic characteristics, the orientation of towns and villages from the bombed cities serves as an ideal natural experiment to identify the mechanisms behind the far-reaching effects of air raids.

The Allied Forces typically launched incendiary attacks against urban industrial areas at altitudes ranging from 800 to 14,000 feet (United States Strategic Bombing Survey 1945–1946). This was the altitude range in which the bombers were visible to the naked eye. Additionally, the Japanese military apprehended incoming bombers with unaided eyes and radars and issued air-raid alerts 15 min before they passed over or reached their target cities (Tokyo Air Raid and War Damage Magazine Editorial Committee 1974, pp. 511–512). Although the purpose of the alert was to inform people about the threat of air raids and encourage them to defend themselves against fire and war, it also incited anxiety and fear. Therefore, not only citizens in bombed cities but also residents of areas untouched by air raids, especially those in the southeast of the cities where the bombers passed overhead, were exposed to the threat of air raids and the associated fear through visual and auditory perception.

# Appendix B  Robustness checks

## B.1  Robustness checks with other data sources

This section presents the tests conducted to verify the robustness of our findings using alternative data sources. The additional test addresses the concern that the data source

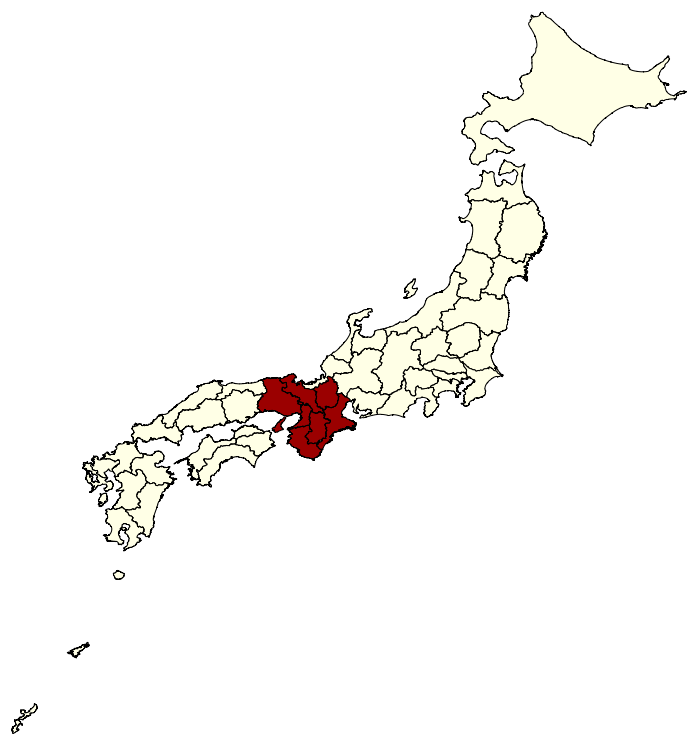

(a) Location of Kinki region

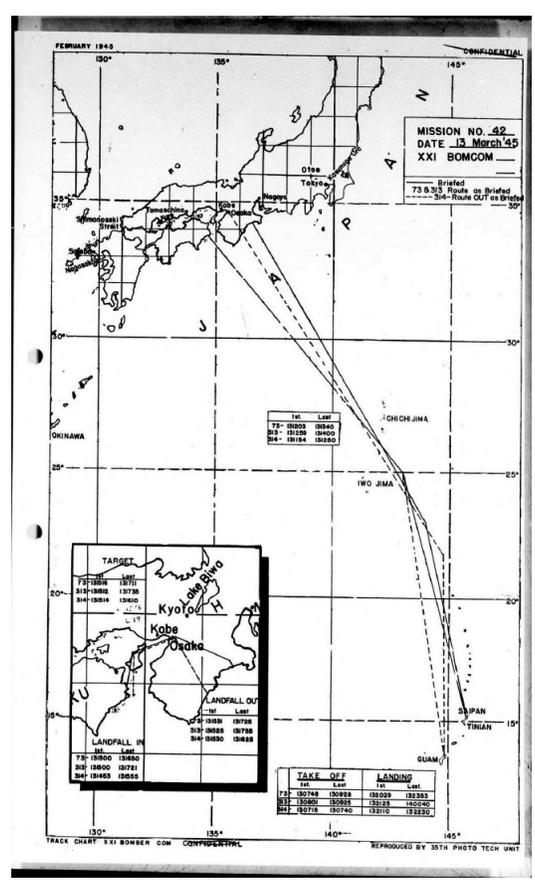

(b) Flight path

Figure A.1: Location of Kinki region and flight path of the Allied bombers

used in our estimates might overestimate the damage from air raids, which could induce a bias in the estimation results . The primary data source for air-raid damages we used was based on reports from the Japanese Economic Stabilization Board. They conducted the survey between 1947 and 1949 because of the need to assess war damages in Japan to reduce reparation payments. Thus, they may have overclaimed the damages from the air raids, even though the damage calculation methodology conformed to the process of the National Wealth Survey conducted by the Statistics Bureau of the Cabinet before the war.

To address this measurement error issue, we used two alternate data sources for the regression analysis. One is the "Final Reports of Records of the U.S. Strategic Bombing

Survey" in United States Strategic Bombing Survey (1945–1946), compiled from USSBS reports: the 20th Air Force Mission Reports, Mission Summaries, Field Orders, Flash Photo Intelligence Reports, and Damage Assessment Reports. After the war ended, they surveyed the damage caused by air raids to evaluate the outcomes of their planned strategic bombings. A notable feature of this data source is its tendency to underestimate the damages compared to our main source. The reason for this underestimation is that these data only cover the damages caused by planned strategic bombings, thereby excluding damages due to other air raids, strafing runs, naval bombardments, and accidental bombings of nontargets. It also excluded potential victims whose deaths were not confirmed because their bodies were not identified.

Another data source is an official document, the "List of Casualties from Air Raids on War-damaged Cities in Japan" in Association of War-Devastated Cities (1962, pp. 71–72), based on a survey report from the Ministry of Construction, possibly for the purpose of reconstruction. This association was a public community organized by municipalities damaged by air raids. Importantly, this data source tends to report more air-raid damages than our primary data, contradicting the possibility that the Economic Stabilization Board immoderately exaggerated the damages.

Table B.1 presents the summary statistics of the death toll from air raids obtained from United States Strategic Bombing Survey (1945–1946) and Association of War-Devastated Cities (1962). Comparing the mean values, the former is 9-28% and the latter is 31-45% larger than our main data. This considerable discrepancy demonstrates the difficulty of calculating war victims, underscoring the importance of estimates based on various data sources. Note that the USSBS data also show a large mean value because this source excludes some cities that suffered less damage.

Table B.2 presents the results estimated using USSBS data, whereas Table B.3 shows the results based on the Association of War-damaged Cities data. The estimation model is the same as the baseline model in Equation (1) in the main text. The estimated coefficients of $Bombing \times Postwar$ in Table B.2 are quite similar to those in Table 2, whereas Table B.3 shows marginally higher values. Overall, these results are consistent with our baseline results, reinforcing the robustness of our findings, regardless of the data source.

Table B.1: Summary statistics of air-raid damages from alternate data sources

| | Death toll from bombing in nearby cities | | | | |
| | Mean | Std. Dev. | Min | Max | Observations |
|---|---|---|---|---|---|
| Panel A: USSBS data | | | | | |
| ≤ 5km | 1,804.48 | 3,103.10 | 4 | 10,293 | 127 |
| ≤ 10km | 2,591.05 | 3,915.19 | 4 | 13,911 | 252 |
| ≤ 15km | 2,788.41 | 4,060.22 | 4 | 13,911 | 435 |
| ≤ 20km | 3,014.32 | 4,163.01 | 4 | 14,651 | 597 |
| ≤ 25km | 3,600.70 | 4,425.44 | 4 | 14,655 | 728 |
| | | | | | |
| Panel B: Association of War-damaged Cities data | | | | | |
| ≤ 5km | 2,582.41 | 3,614.16 | 8 | 12,264 | 115 |
| ≤ 10km | 3,696.18 | 4,539.75 | 38 | 18,066 | 219 |
| ≤ 15km | 4,050.56 | 4,882.48 | 8 | 18,421 | 374 |
| ≤ 20km | 4,181.03 | 5,040.52 | 8 | 20,305 | 545 |
| ≤ 25km | 4,718.47 | 5,379.24 | 8 | 20,305 | 711 |

Notes: Only non-zero values are reported. Although it appears anomalous that Panel B has a minimum value of 38 only within a 10 km radius, this is attributed to the geographical distribution of municipalities. Fuse City recorded eight fatalities due to air raids, which is the minimum value in other cases. However, there was no town or village where Fuse was the only damaged city within 10 km.
Sources: United States Strategic Bombing Survey (1945–1946) and Association of War-Devastated Cities (1962).

Table B.2: Effects of bombing estimated with USSBS data

| | (1) | (2) | (3) | (4) | (5) |
|---|---|---|---|---|---|
| *Bombing × Postwar* | 0.283*** | 0.284*** | 0.326*** | 0.304*** | 0.290*** |
| | (0.051) | (0.051) | (0.059) | (0.060) | (0.108) |
| *Sex ratio* | | 1.284 | 2.349 | 2.219 | 2.231 |
| | | (2.771) | (2.847) | (2.851) | (2.840) |
| ln*Population* | | | −2.258 | −2.593* | −2.565* |
| | | | (1.503) | (1.540) | (1.539) |
| *Proximity to city × Postwar* | | | | 0.540 | |
| | | | | (0.357) | |
| ln(*City population*) | | | | | −0.069 |
| | | | | | (1.164) |
| ln(*City population*) × *Postwar* | | | | | 0.040 |
| | | | | | (0.032) |
| Town and village fixed effects | Yes | Yes | Yes | Yes | Yes |
| Postwar fixed effect | Yes | Yes | Yes | Yes | Yes |
| Observations | 2,874 | 2,874 | 2,874 | 2,874 | 2,874 |

Notes: The dependent variable is the crude birth rate (per mil). The independent variable *Bombing* is the number of deaths (in the thousands) due to the bombing in cities within 15 km. The *Sex ratio* is the male-to-female sex ratio, the ln*Population* is the log of population, the *Proximity to city* is the dummy variable taking one if there were cities within 15 km, and the *City population* is the inverse hyperbolic sine of the total population in cities within 15 km. ***, **, and * represent statistical significance at the 1%, 5%, and 10% levels, respectively. Standard errors clustered at the town–village level are in parentheses.

## B.2 Verification using other measures of bombing damage

We applied the death toll as the primary measure of bombing damage for the regression analyses, given its unambiguous criterion for calculation in the survey. However, air raids

Table B.3: Effects of bombing estimated with Association of War-damaged Cities data

| | (1) | (2) | (3) | (4) | (5) |
|---|---|---|---|---|---|
| *Bombing×Postwar* | 0.226*** | 0.226*** | 0.261*** | 0.241*** | 0.225*** |
| | (0.042) | (0.043) | (0.050) | (0.051) | (0.092) |
| *Sex ratio* | | 1.241 | 2.245 | 2.110 | 2.130 |
| | | (2.771) | (2.846) | (2.850) | (2.841) |
| ln*Population* | | | −2.139 | −2.456 | −2.436 |
| | | | (1.502) | (1.538) | (1.538) |
| *Proximity to city×Postwar* | | | | 0.537 | |
| | | | | (0.357) | |
| ln(*City population*) | | | | | −0.131 |
| | | | | | (1.158) |
| ln(*City population*)×*Postwar* | | | | | 0.042 |
| | | | | | (0.032) |
| Town and village fixed effects | Yes | Yes | Yes | Yes | Yes |
| Postwar fixed effect | Yes | Yes | Yes | Yes | Yes |
| Observations | 2,874 | 2,874 | 2,874 | 2,874 | 2,874 |

Notes: The dependent variable is the crude birth rate (per mil). The independent variable *Bombing* is the number of deaths (in the thousands) due to the bombing in cities within 15 km. The *Sex ratio* is the male-to-female sex ratio, the ln*Population* is the log of population, the *Proximity to city* is the dummy variable taking one if there were cities within 15km, and the *City population* is the inverse hyperbolic sine of the total population in cities within 15 km. ***, **, and * represent statistical significance at the 1%, 5%, and 10% levels, respectively. Standard errors clustered at the town–village level are in parentheses.

do not always lead to death. Some people may have been fortunate enough to avoid death despite the severe aerial bombardment. Thus, it is possible that the number of deaths reflects an inappropriate scale of the air raids. To confirm that our findings are robust regardless of the variables representing bombing scales, we conducted sensitivity tests using measures other than the death toll: the number of injured people, casualties (sum of dead and injured), and damaged buildings (total of completely destroyed, half-destroyed, completely burned, and half-burned).

Table B.4 reports the estimation results of the bombing on the postwar crude birth rate using different measures of air-raid scale. All specifications include the sex ratio, population, proximity dummy, and fixed effects. The key independent variables were standardized to have a mean of zero and variance of one for ease of comparison. As a benchmark, Column (1) presents the death toll result, which is essentially identical to the baseline result in Column (4) of Table 2. Column (2) shows the estimated coefficient of the number of injured people, whereas Column (3) indicates the sum of dead and injured people. The results using the number of damaged buildings are shown in Column (4). These damage measures interact with the postwar dummy variable in the estimating

Table B.4: Estimation results of alternate measures of bombing damage

| Damage measure | (1) | (2) | (3) | (4) |
|---|---|---|---|---|
| Dead | 0.690*** | | | |
| | (0.156) | | | |
| Injured | | 0.808*** | | |
| | | (0.148) | | |
| Dead and injured | | | 0.778*** | |
| | | | (0.150) | |
| Building | | | | 0.813*** |
| | | | | (0.148) |
| All control variables | Yes | Yes | Yes | Yes |
| Town and village fixed effects | Yes | Yes | Yes | Yes |
| Postwar fixed effect | Yes | Yes | Yes | Yes |
| Observations | 2,874 | 2,874 | 2,874 | 2,874 |

Notes: The dependent variable is the crude birth rate (per mil). The key independent variables are damage measures due to air raids in cities within 15 km. The control variables are the male-to-female sex ratio, log of population, and the proximity dummy variable taking one if there were cities within 15 km. ***, **, and * represent statistical significance at the 1%, 5%, and 10% levels, respectively. Standard errors clustered at the town–village level are in parentheses.

equations.

The estimated coefficients were positive and statistically significant across all measures of the air-raid scale, consistent with the baseline results. The minimal magnitude of the effect size in Column (1) suggests that our main estimates using the conservative death toll led to conservative results. Aerial bombing could have had far-reaching effects on fertility, even without causing deaths.

We also estimated the effects of the bombings on the crude birth rate by damage intensity cluster measured using different indicators. Figure B.1 shows the results. The scale of damage from the air raids gradually increased from the first to the fifth cluster. Although there are nuanced distinctions, all the measures consistently indicate the same overall trend, as shown in Fig. 5: Smaller air raids had negative effects, whereas larger raids had positive effects. These results strengthened the robustness of our findings.

## B.3 Estimates using subsample

While historical evidence shows that citizens dispersedly evacuated to their relatives, one might consider that those exposed to air raids tended to evacuate to rural areas far from the cities and settled there. This concern implies that our regression analyses fail to capture the far-reaching effects of bombing owing to the influence of direct air raids in

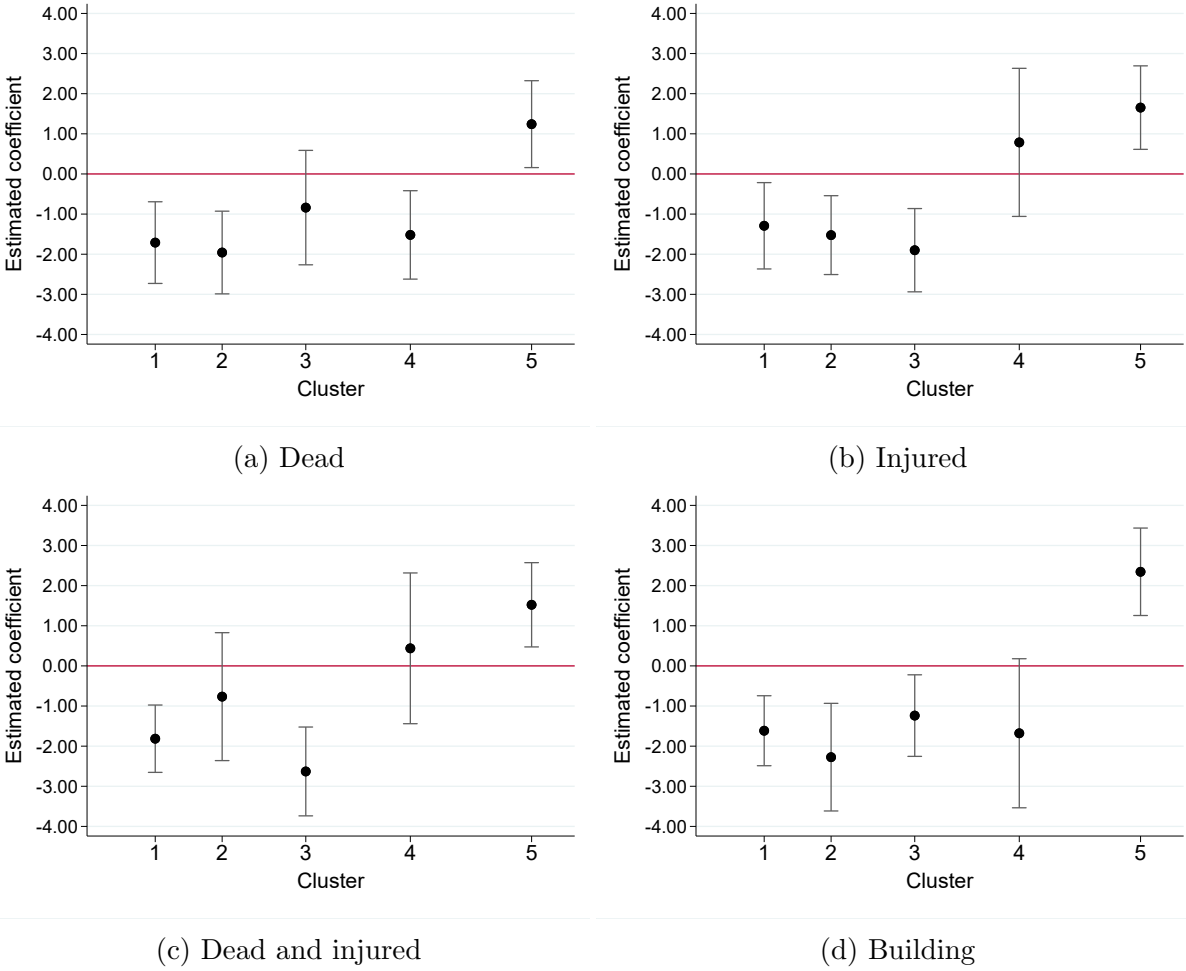

(a) Dead

(b) Injured

(c) Dead and injured

(d) Building

Figure B.1: Effects of bombing by damage size based on different measures

Notes: The dotted and solid lines indicate the point estimates and their 90% confidence intervals, respectively. The dependent variable is the crude birth rate (per mil). The key independent variables are the dummy variables representing each cluster of bombing damage intensity within 15 km. The control variables are the male-to-female sex ratio, log of population, and dummy variable representing proximity to a city. The town and village fixed effects and postwar dummy variable are also included. Robust standard error is clustered at the town–village level.

the reference group. Furthermore, the characteristics of towns and villages could have varied according to their geographical proximity to a city, which might have disturbed our estimates. Thus, we performed analyses using a subsample of municipalities within 15 kilometers of a city instead of including the proximity dummy in the estimating equation. This approach allowed us to estimate far-reaching effects, excluding towns and villages that potentially had fertility characteristics different from those in the treatment group.

Figure B.2 displays the results of the bombing effects on the post-war crude birth rate by damage intensity cluster, estimated from observations only within 15 km of the cities. The estimation specifications are presented in Figs. B.2a and B.2b are identical to those

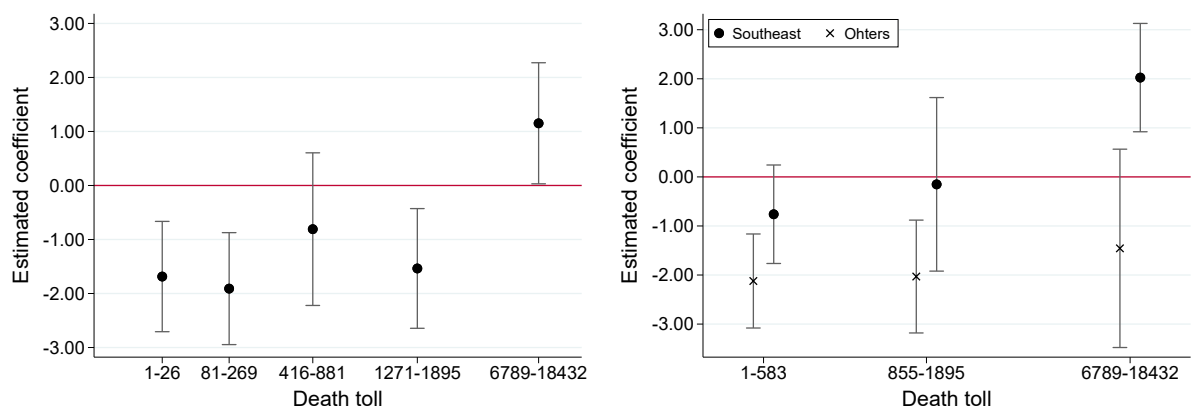

(a) Effects by damage cluster

(b) Effects by damage cluster and direction

Figure B.2: Effects of bombing by damage size estimated from subsample

Notes: The dot and cross marks denote the point estimates, whereas the solid lines indicate their 90% confidence intervals. The dependent variable is the crude birth rate (per mil). The control variables are the male-to-female sex ratio, and log of population, in addition to the town and village fixed effects. The number of observations are 1,696, comprising 848 towns and villages within 15 km of the cities. Robust standard error is clustered at the town–village level.

in Figs. 5 and 7 in the main text, respectively. Evidently, the estimation results remain unchanged if we use the subsample. The stability of the results supports the validity and robustness of the main estimates.



\end{document}